%% file: H0_with_zdm.tex
\documentclass[fleqn,usenatbib,useAMS]{mnras}

\usepackage[percent]{overpic}
\usepackage{newtxtext,newtxmath}

\usepackage[T1]{fontenc}
\usepackage{ae,aecompl}
\usepackage{multirow}

\usepackage{graphicx}	
\usepackage{amsmath}	

\newcommand{\james}{J22a}  
\newcommand{\effect}{the cliff effect}
\newcommand{\result}{\ensuremath{73_{-8}^{+12}}}
\newcommand{\icspaper}{Shannon et al.\ (in prep.)}
\newcommand{\spirals}{Deller et al.\ (in prep.)}
\newcommand{\bhandariinprep}{Bhandari et al.\ (in prep.)}

\newcommand{\Hnot}{\ensuremath{{H_{0}}}}
\newcommand{\hubbunit}{\ensuremath{\rm km \, s^{-1} \, Mpc^{-1}}}
\newcommand{\omegabhh}{\ensuremath{\Omega_b h^2}}
\newcommand{\omegabh}{\ensuremath{\Omega_b h}}
\newcommand{\omegab}{\ensuremath{\Omega_b}}
\newcommand{\omegam}{\ensuremath{\Omega_{\rm m}}}

\newcommand{\pccc}{\ensuremath{{\rm pc}\,{\rm cm}^{-3}}}

\newcommand{\dmunits}{\ensuremath{{\rm pc \, cm^{-3}}}}
\newcommand{\dmcosmic}{\ensuremath{{\rm DM}_{\rm cosmic}}}
\newcommand{\dmacosmic}{\ensuremath{\langle {\rm DM}_{\rm cosmic} \rangle}}
\newcommand{\dmfrb}{\ensuremath{{\rm DM}_{\rm FRB}}}
\newcommand{\dmigm}{\ensuremath{{\rm DM}_{\rm IGM}}}
\newcommand{\dmhost}{\ensuremath{{\rm DM}_{\rm host}}}

\newcommand{\dmhalo}{\ensuremath{{\rm DM}_{\rm halo}}}
\newcommand{\dmism}{\ensuremath{{\rm DM}_{\rm ISM}}}
\newcommand{\dmeg}{\ensuremath{{\rm DM}_{\rm EG}}}
\newcommand{\dmlocal}{\ensuremath{{\rm DM}_{\rm local}}}
\newcommand{\vdmhost}{\ensuremath{186_{-48}^{+59}}}

\newcommand{\fe}{CRAFT/FE}
\newcommand{\mb}{Parkes/Mb}
\newcommand{\ics}{CRAFT/ICS}
\newcommand{\icshigh}{CRAFT/ICS \ensuremath{1.6\,{\rm GHz}}}
\newcommand{\icsmid}{CRAFT/ICS \ensuremath{1.3\,{\rm GHz}}}
\newcommand{\icslow}{CRAFT/ICS \ensuremath{900\,{\rm MHz}}}

\newcommand{\figref}[1]{Figure~\ref{#1}} 
\newcommand{\secref}[1]{\S\ref{#1}}
\newcommand{\tabref}[1]{Table~\ref{#1}}

\newcommand{\emin}{\ensuremath{E_{\rm min}}}
\newcommand{\emax}{\ensuremath{E_{\rm max}}}
\newcommand{\Eth}{\ensuremath{E_{\rm th}}}
\newcommand{\lemax}{$\log_{10} E_{\rm max}$}
\newcommand{\lemin}{$\log_{10} E_{\rm min}$}

\newcommand{\pdm}{\ensuremath{P({\rm DM})}}
\newcommand{\pzdm}{\ensuremath{P(z,{\rm DM})}}
\newcommand{\pszdm}{\ensuremath{P(z,{\rm DM},\sss)}}

\newcommand{\pdmgz}{\ensuremath{P({\rm DM}|z)}}
\newcommand{\pzgdm}{\ensuremath{P(z|{\rm DM})}}
\newcommand{\pzs}{\ensuremath{P(z,\sss)}}
\newcommand{\pn}{\ensuremath{P(\Nfrb)}}

\newcommand{\sss}{\ensuremath{\rm SNR}}  
\newcommand{\snr}{\ensuremath{{\rm SNR}}}
\newcommand{\snrth}{\ensuremath{{\rm SNR}_{\rm th}}}
\newcommand{\Nfrb}{\ensuremath{N_{\rm FRB}}}

\newcommand{\muhost}{\ensuremath{\mu_{\rm host}}}
\newcommand{\sigmahost}{\ensuremath{\sigma_{\rm host}}}

\newcommand{\sfrn}{\ensuremath{n_{\rm sfr}}}

\newcommand{\pox}{\ensuremath{P(O|x)}}

\newcommand{\fdz}{\ensuremath{f_d(z)}} 
\newcommand{\fdzo}{\ensuremath{f_d(z=0)}} 
\newcommand{\fdzn}{\ensuremath{f_d(z=1)}} 

\title[\Hnot\ with FRBs]{A measurement of Hubble’s Constant using Fast Radio Bursts} 

\author[James et al.]{
C.W.~James,${^{1}}$\thanks{E-mail: clancy.james@curtin.edu.au}
E.M. Ghosh,${^{2}}$\thanks{E-mail: esanmoulig@gmail.com}
J.X.~Prochaska,${^{3,4}}$\thanks{E-mail: xavier@ucolick.org} 
K.W. Bannister,${^{5}}$\newauthor 
S. Bhandari,${^{6,7,8,5}}$ 
C.K. Day,${^{9,10}}$ 
A.T. Deller,${^{10}}$ 
M. Glowacki,${^{1}}$
A.C. Gordon,${^{11}}$\newauthor 
K.E. Heintz,${^{12}}$ 
L. Marnoch,${^{13,5,14}}$ 
S.D. Ryder,${^{13,14}}$ 
D.R. Scott,${^{1}}$
R.M. Shannon,${^{10}}$\newauthor 
N. Tejos$^{15}$ 
\\ 
$^{1}$International Centre for Radio Astronomy Research, Curtin University, Bentley, WA 6102, Australia \\
$^{2}$Indian Institute of Science Education and Research Mohali, Knowledge City, Sector 81, SAS Nagar, Manauli PO 140306, India\\
$^{3}$Department of Astronomy and Astrophysics, University of California, Santa Cruz, CA 95064, USA\\
$^{4}$Kavli Institute for the Physics and Mathematics of the Universe, 5-1-5 Kashiwanoha, Kashiwa 277-8583, Japan\\
$^5$CSIRO, Space and Astronomy, PO Box 76, Epping NSW 1710 Australia \\
$^6$ASTRON, Netherlands Institute for Radio Astronomy, Oude Hoogeveensedijk 4, 7991 PD Dwingeloo, The Netherlands \\
$^7$Joint institute for VLBI ERIC, Oude Hoogeveensedijk 4, 7991 PD Dwingeloo, The Netherlands \\
$^8$Anton Pannekoek Institute for Astronomy, University of Amsterdam, Science Park 904, 1098 XH, Amsterdam, The Netherlands \\
$^9$ Department of Physics, McGill University, Montreal, Quebec H3A 2T8, Canada\\
$^10$Centre for Astrophysics and Supercomputing, Swinburne University of Technology, P.O. Box 218, Hawthorn, VIC 3122, Australia \\
$^{11}$Center for Interdisciplinary Exploration and Research in Astrophysics and Department of Physics and Astronomy,\\ Northwestern University, 2145 Sheridan Road, Evanston, IL 60208-3112, USA \\
$^{12}$Cosmic Dawn Center (DAWN), Niels Bohr Institute, University of Copenhagen, Jagtvej 128, 2100 Copenhagen {\o}, Denmark \\
$^{13}$School of Mathematical and Physical Sciences, Macquarie University, NSW 2109, Australia \\
$^{14}$Astronomy, Astrophysics and Astrophotonics Research Centre, Macquarie University, Sydney, NSW 2109, Australia \\
$^{15}$ Instituto de F\'isica, Pontificia Universidad Cat\'olica de Valpara\'iso, Casilla 4059, Valpara\'iso, Chile
}

\date{Accepted XXX. Received YYY; in original form ZZZ}

\pubyear{2021}

\begin{document}

\label{firstpage}
\pagerange{\pageref{firstpage}--\pageref{lastpage}}
\maketitle

\begin{abstract}
We constrain the Hubble constant \Hnot\ using Fast Radio Burst (FRB) observations from the Australian Square Kilometre Array Pathfinder (ASKAP) and Murriyang (Parkes) radio telescopes.
    We use the redshift-dispersion measure (`Macquart') relationship, accounting for the intrinsic luminosity function, cosmological gas distribution, population evolution, host galaxy contributions to the dispersion measure (\dmhost), and observational biases due to burst duration and telescope beamshape. Using an updated sample of 16 ASKAP FRBs detected by the Commensal Real-time ASKAP Fast Transients (CRAFT) Survey and localised to their host galaxies, and 60 unlocalised FRBs from Parkes and ASKAP, our best-fitting value of \Hnot\ is calculated to be \result\,\hubbunit.
    Uncertainties in FRB energetics and \dmhost\ produce larger uncertainties in the inferred value of \Hnot\ compared to previous FRB-based estimates. Using a prior on \Hnot\ covering the 67--74\,\hubbunit\ range, we estimate a median $\dmhost = \vdmhost \dmunits$, exceeding previous estimates. We confirm that the FRB population evolves with redshift similarly to the star-formation rate.
    We use a Schechter luminosity function to constrain the maximum FRB energy to be \lemax$=41.26_{-0.22}^{+0.27}$\,erg assuming a characteristic FRB emission bandwidth of 1 GHz at 1.3 GHz, and the cumulative luminosity index to be $\gamma=-0.95_{-0.15}^{+0.18}$.
    We demonstrate with a sample of 100 mock FRBs that \Hnot\ can be measured with an uncertainty of $\pm 2.5$\,\hubbunit, demonstrating the potential for clarifying the Hubble tension with an upgraded ASKAP FRB search system. Last, we explore a range of sample and selection biases that affect FRB analyses.

\end{abstract}

\begin{keywords}
fast radio bursts -- cosmological parameters
\end{keywords}

\newpage

\section{Introduction}

Fast radio bursts (FRBs) are millisecond-duration pulses of radio emission  observed at frequencies from $\sim$ 100 MHz up to a $\sim8$\,GHz now known to originate at cosmological distances  \citep{Lorimer2007,Shannonetal2018,Gajjar2018,CHIME_catalog1_2021,Pleunis2021}. Their progenitors and burst production mechanism are as yet unknown and many progenitor models have been proposed \citep{Platts2018}. FRBs have also been observed to repeat \citep[e.g.][]{spitler2016repeating}, with two showing cyclical phases of irregular activity \citep{2020_121102_periodicity1,Chime2020repetition}.  There is evidence that FRBs come from more than one source class \citep[e.g.][]{CHIME_morphology_2021}, although it is also possible that apparent morphological differences in the time--frequency properties of the FRB population can be produced by a single progenitor \citep{Hewitt2021}.

Despite uncertainties as to their origins, FRBs have the potential to act as excellent cosmological probes to trace the ionised gas and magnetic fields in galaxy halos, large-scale structure, and the intergalactic medium 
\citep{McQuinn2014,MasuiSigurdson2015,ProchaskaZheng2019,Madhavacheril2019, CalebHelium2019, lee+2022}.
This is because the radio pulse from the burst is dispersed while travelling through the ionized intergalactic medium, with the total inferred dispersion measure (DM) being a powerful probe of the column density
of ionised electrons along the line of sight.
Recently, localised FRBs have been used to 
resolve the `missing baryons problem'  \citet{Macquart2020}, where the probability distribution of observed DM given the 
redshift $z$ of identified FRB host 
galaxies is analysed to constrain the total baryon density of the Universe and the degree of galactic baryon feedback.

Additionally, FRBs can be used to measure the value of the Hubble constant.
The cosmic expansion rate $\dot{a}(t)$
can be expressed in terms of the Hubble parameter $H(z) = \dot{a}(t)/a(t)$. In a flat $\Lambda$CDM cosmology, $H(z)$ (sometimes written as $\Hnot = 100 h\,$\hubbunit) can be expressed as $H(z) = \Hnot\sqrt{\Omega_{\Lambda}+\omegam(1+z)^3}$
where \Hnot\ is the Hubble constant, $\Omega_{\Lambda}$ is the vacuum energy density fraction, and $\omegam$ is the
matter density fraction, at $z=0$. The value of 
\Hnot\ characterises the expansion rate of the Universe at the present time, and determines its absolute distance scale. There has been remarkable progress in improving the accuracy of \Hnot\ measurements from local-Universe measurements, with the  10\% uncertainty from the \emph{Hubble Space Telescope} \citep{Freedman_2001} improving to less than 1\%  more recently
\citep[e.g.][]{Riess_2016,Suyu_2017}. 
However, there exists a $\sim4\sigma$ tension between measurements of the Hubble constant inferred from Planck observations of the Cosmic Microwave Background (CMB) which is $\Hnot=67.4\pm 0.5 \hubbunit$
\citep{PlanckCosmology2018}, and those made from calibrating standard candles such as the expanded sample of local type Ia supernovae (SNe Ia) calibrated by the distance ladder 
\citep[\Hnot$=73.04\pm 1.04$\,\hubbunit;][]{Riess2021}.
Thus far,  studies of observational biases and systematic uncertainties
have not alleviated this tension, motivating solutions that include involving early or dynamical dark energy, neutrino interactions, interacting cosmologies, primordial magnetic fields, or modified gravity in our understanding of the $\Lambda$CDM model  --- see \citet{SNOWMASScosmology} for a recent review. Therefore an independent and robust method of measuring \Hnot\
would be a welcome addition to the tools of physical cosmology.

Analysis of FRB observations 
offer such an independent and local ($z<1$) test.
Two direct observations of FRBs -- DM and the signal-to-noise ratio (\snr) --
and one inferred property based on host galaxy associations (redshift, $z$)
provide the set of constraints on \Hnot.
There are two, largely independent constraints at work.
One is effectively a standard candle analysis.
To the extent that the FRB energetics are independent
of redshift, an ansatz,
the \snr\ dependence with redshift is sensitive to \Hnot.
This constraint, however, is highly degenerate with the (unknown) intrinsic distribution of FRB energies. The other constraint is set by the cosmic contribution
to the FRB DM (\dmfrb), referred to as \dmcosmic.
The average value of \dmcosmic, 
$\dmacosmic \propto \Omega_b H_0$ and 
to the extent that $\Omega_b H_0^2$ is
precisely measured by CMB and  Big Bang Nucleosynthesis analysis
\citep{PlanckCosmology2018,Mossa2020_Deuterium}, this implies
$\dmacosmic \propto H_0^{-1}$.
Therefore, the distribution of \dmfrb\ and redshifts
offer a direct constraint on \Hnot.

To leverage FRBs, one requires 
a detailed study of the observed distribution 
of FRBs in \sss, $z$, and DM space, 
\pszdm. 
\citet[][hereafter \james]{James2022Meth} have developed an advanced model of FRB observations using the Australian Square Kilometre Array Pathfinder (ASKAP) and Murriyang (Parkes) radio telescope data, accounting for observational biases (due to burst temporal width, DM, and the exact telescope beamshape) to assess \pszdm. They estimated that unlocalised ASKAP FRBs arise from z < 0.5, with between a third and a half within z < 0.1, and find that above a certain DM, observational biases cause the observed Macquart (DM-–z) relation to become inverted, implying that the highest-DM events detected in the unlocalised Parkes and ASKAP samples are unlikely to be the most distant. Thus analyses assuming a one-to-one z--DM relationship may lead to biased results, particularly in this high-DM regime.

In this paper, we extend the model developed by \james\ 
to constrain \Hnot. 
The modelling of \pszdm\ is described in 
\secref{sec:p(DM,z)}, along with the distribution of \dmhost, \dmcosmic, $\Phi$ (rate of FRBs per comoving volume) and the FRB luminosity function. The detection efficiency and beamshape sensitivity of the surveys are also taken into consideration to calculate the final distribution of FRBs in ($z$,DM) space. Our sensitivity to \Hnot\ is described in \secref{sec:explore}. In \secref{sec:sample} the properties of the FRB sample data used from Parkes and ASKAP radio telescopes is described, where we include a total of 16 ASKAP FRBs localised by the Commensal Real-time ASKAP Fast Transients (CRAFT) Survey. In \secref{sec:results} we perform a Bayesian analysis to determine the best-fitting value of \Hnot\ given our dataset. In \secref{sec:craco_mc_results} we test the validity of our model by creating mock sample surveys using Monte Carlo simulations and checking whether the best-fitting value of \Hnot\ obtained is close to the truth value of \Hnot\ at which the samples are created. \secref{sec:discussion} contains a discussion on these results and on future prospects of precision cosmology using an extended FRB dataset.

\section{Forward Modelling the \pszdm\ distribution 
of FRBs}
\label{sec:p(DM,z)}

Our study is based on comparing three observables
related to FRBs to a forward model:
 (i) the fast radio burst dispersion measure, \dmfrb;
 (ii) the \snr\ of the pulse relative
 to the survey threshold, $s$;
 and
 (iii) when available, the redshift $z$ of the FRB 
 determined by a high probability association to
 its host galaxy.
Details on these quantities and the observational
sample are presented in the following section.

The methodology for our forward model was 
introduced in \james\ and applied to several 
surveys of FRBs.
In this manuscript, we present an extension of
their model to analyze \Hnot.
We offer a brief summary of the model here,
mainly emphasizing the aspects that vary with
\Hnot, and also
detail any updates or changes to the model.

\subsection{Dispersion Measure}
\label{sec:DM}

The DM of a radio pulse is 
the integrated number density of free electrons 
along the propagation path. This causes a delay between the arrival times of different pulse
frequencies $\nu$. 
As an integral measure, \dmfrb\ includes contributions
from several components which we model separately.
\dmfrb\ is divided into an ‘extra-galactic’ 
contribution,
\dmeg\ and a contribution from the ‘local’ Universe \dmlocal:
\begin{equation}
\dmfrb = \dmeg(z) + \dmlocal, \label{eq:dmfrb}
\end{equation}
where 
\begin{equation}
\dmeg(z) \equiv \dmcosmic(z) + \frac{\dmhost}{1+z}\ \; ,\label{eq:dmeg}
\end{equation}
and
\begin{equation}
\dmlocal \equiv \dmism(l,b) + \dmhalo \; . \label{eq:dmlocal}
\end{equation}
which includes respective contributions from 
the Milky Way’s interstellar medium (ISM, \dmism), 
its Galactic halo (\dmhalo), 
the cosmological distribution of ionised gas
(\dmcosmic), and the FRB host (\dmhost). The latter incorporates the host galaxy halo, ISM, and any contribution from the small-scale environment surrounding the FRB progenitor.
The NE2001 model \citep{CordesLazio01} is used to estimate $\dmism(l,b)$ as a function of Galactic coordinates $(l,b)$,
while \dmhalo\ is set to be 
$50\,\dmunits$ based on estimates
from other works \citep{ProchaskaZheng2019,KeatingPen2020,Platts2020}.  
In practice, the \dmhalo\ value is 
largely degenerate
with our model for \dmhost\ (but see our discussion in \secref{sec:discussion}).
For \dmhost, we adopt the log-normal probability
distribution of \james, with parameters \muhost\ and \sigmahost.

The only significant change to the \james\ prescription
for DM is on the cosmological contribution
\dmcosmic\ which has an explicit dependence
on \Hnot.
Adopting the cosmological paradigm of a flat Universe with matter
and dark energy, the average value of 
\dmcosmic\ is calculated as \citep{Inoue2004}:
\begin{eqnarray}
\dmacosmic  & = & \int\limits_0^z \frac{c \bar{n}_e(z^\prime) \, dz^\prime}{H_0 (1+z^\prime)^2 {\rm E}(z)} \label{eq:H0},\\
{\rm with} \; {\rm E}(z) & = & \sqrt{\omegam(1+z^\prime)^3 + \Omega_\Lambda}, \label{eq:Ez}
\end{eqnarray}
with $\bar n_e$ the mean density of electrons,
\begin{eqnarray}
\bar{n}_e = \fdz\rho_{b}(z)m_{p}^
{-1} \chi_{e} \label{eq:nebar}\\
= \fdz\rho_{b}(z)m_{p}^
{-1} (1 - {\rm Y}_{\rm He}/2)
\end{eqnarray} 
with $\chi_{e}= Y_{\rm{H}}+Y_{\rm{He}}/2 \approx 1-Y_{\rm{He}}/2$ 
calculated from the primordial hydrogen and helium mass fraction $Y_{\rm{H}}$ and $Y_{\rm{He}}$. 
This is found to be 0.25 (assumed doubly ionized helium) to high precision by CMB measurements \citep{PlanckCosmology2018}; current best estimates are $0.2453 \pm 0.0034$ \citep{Aver2021}.
Furthermore, $m_{p}$ is the proton mass,
$\fdz$ is the fraction of cosmic baryons in diffuse ionized gas, and $\rho_{b}$ is the mass
density of baryons defined as 
\begin{eqnarray}
\rho_{b}(z)= \Omega_{b}\rho_{c,0}(1 + z)^3, \label{eq:rhob} 
\end{eqnarray}
with $\rho_{c,0}$ the critical density and
$\Omega_b$ the baryon density parameter.

Throughout the analysis we adopt the 
\citet{PlanckCosmology2018} set of cosmological
parameters except for \Hnot\ and $\Omega_b$ (uncertainties in the former are sub-dominant compared to other sources --- see \secref{sec:discussion}).
Because $\rho_c \equiv 3 H_0^2 / 8 \pi G$, 
\eqref{eq:H0}, \eqref{eq:nebar} and \eqref{eq:rhob} imply
\begin{equation}
\dmacosmic\ \propto \bar n_e H_0^{-1} \propto \Omega_b H_0 \; .
\label{eqn:invH}
\end{equation}

Two complementary methods -- (1) deuterium to hydrogen
measurements coupled with BBN
theory and (2) CMB measurements and analysis --
have constrained $\Omega_b H_0^2$ to 
$\approx 1\%$ precision \citep{cooke2018,mossa2020}.
Therefore, we consider
$\Omega_b (H_0/100)^2$ a 
fixed constant of 0.02242 \citep{PlanckCosmology2018}.\footnote{The latest measurement, from primordial deuterium abundances, is $\omegabhh=0.02233 \pm 0.00036$ \citep{Mossa2020_Deuterium}.
In future works, we will allow for the small
uncertainty in $\Omega_b H_0^2$,
but it has a negligible contribution
to the current results.
}
Thus when we vary \Hnot, $\Omega_b$ is
adjusted accordingly. This yields
\begin{equation}
\dmacosmic\ \propto H_0^{-1} \; ,
\end{equation}
which we explore further in
$\S$~\ref{sec:explore}.

Regarding \fdz,  we adopt the approach
derived in \citet{Macquart2020} which combines
estimates for the Universe's baryonic components 
that do not contribute to \dmcosmic\
(e.g.\ stars, stellar remnants, neutral gas).
Current estimates yield 
$f_d (z=0) = 0.844$ with uncertainties of
a few percent (dominated by uncertainties in the
initial mass function of stars). Evidence suggests an evolving \fdz\ \citep{Lemos2022}, and we use the implementation in \citet{frb} to describe this.
For the current study, the uncertainty in
\fdz\ is unimportant, yet it may become a limiting
systematic in the era of many thousands of
well-localized FRBs, as we discuss further in \secref{sec:uncertainties}.

\subsection{Rate of FRBs}

Our model of the FRB population is primarily described in \james, much of which is in-turn based on \citet{Macquart2018b}. Here, we describe only modifications to that model.

\subsubsection{Population Evolution}
\label{sec:population_evolution}

We model the rate of FRBs per comoving volume $\Phi(z)$ as a function of redshift, specifically to some power \sfrn\ of the star formation rate according to \citet{Macquart2018b},
\begin{eqnarray}
\Phi(z) & = & \frac{\Phi_0}{1+z} \left( \frac{{\rm SFR}(z)}{{\rm SFR}(0)} \right)^{\sfrn}, \label{eq:phiz}
\end{eqnarray}
and SFR$(z)$ from \citet{MadauDickinson_SFR},
\begin{eqnarray}
{\rm SFR}(z) & = & 1.0025738 \frac{(1+z)^{2.7}}{1 + \left(\frac{1+z}{2.9}\right)^{5.6}}. \label{eq:sfr_n}
\end{eqnarray}
The motivation for this formalism is to allow a smooth scaling between no source evolution ($\sfrn=0$), evolution with the SFR ($\sfrn=1$), and a more-peaked scenario similar to AGN evolution ($\sfrn\sim2$).
The total FRB rate in a given redshift interval $dz$ and sky area $d\Omega$ will also be proportional to the total comoving volume $dV$,
\begin{eqnarray}
\frac{dV}{d\Omega dz} & = & D_H \frac{(1+z)^2 D_A^2(z)}{E(z)}, \label{eq:comoving_volume}
\end{eqnarray}
which depends on the angular diameter distance $D_A$, as well as Hubble distance $D_H=c/H_0$. Thus for a higher value of Hubble's constant, the rate of FRBs in a comoving volume $dV$ will be lower, assuming the SFR remains constant. 

\subsubsection{FRB Luminosity Function}

In \james, we modeled the 
FRB luminosity function by a simple power-law distribution $p(E) \propto E^\gamma$ bounded
by a minimum and maximum energy (\emin,\emax). 
We use `burst energy' as the isotropic equivalent energy at $1.3$\,GHz (i.e.\ beaming is ignored), and use an effective bandwidth of 1\,GHz when converting between spectral and bolometric luminosity. 
While we find
this simple distribution is still a sufficient
description of the observational data,
we now adopt an upper incomplete Gamma function as our cumulative energy distribution,
\begin{eqnarray}
p(E > \Eth) = \int_{\Eth}^{\infty} (E/\emax)^\gamma \exp(-E/\emax) dE, \label{eq:schechter}
\end{eqnarray}
the derivative of which is often termed the `Schechter' function. This eliminates numerical artefacts in the
analysis of \Hnot\ that can arise due to the infinitely
sharp cutoff in the power-law at 
$E>\emax$.

Although \citet{Li2021_FAST_121102} find a minimum burst energy for FRB~20121102, our analysis in \james\ showed no evidence of a minimum value of burst energy for the FRB population as a whole, so we set the value of $\emin = 10^{30} \, \rm erg$,
which is several orders of magnitude below the minimum burst energy of any FRB detected. 

Individual FRBs show detailed structure in both the time and frequency domain \citep{CHIME_morphology_2021}, which in the case of repeaters, is also highly time-variable \citep{Hessels2019}. As we have discussed in \james, this introduces ambiguities in modelling their spectral properties. We choose to use the `rate interpretation' for FRB spectral behaviour, where FRBs are narrow in bandwidth, and have a frequency dependent rate ($\Phi(\nu) \propto \nu^\alpha$). This provides an equally good description of FRB properties to the more usual `spectral index' interpretation, in which FRBs have fluences that scale with frequency; and it is computationally much faster to implement. 

\subsubsection{Scattering}
\label{sec:scattering}

The FRB width model used in \james\ modelled the total width distribution of FRBs (i.e.\ including intrinsic width $w_i$ and scattering $w_s$) as a log-normal, with mean $\log \mu_w [{\rm ms}]=1.70$ and $\log \sigma_w = 0.73$. This was based on the fit to observed \fe\ (CRAFT Fly's Eye) and \mb{} (Parkes multibeam) FRBs from \citet{Arcus2020}, and accounted for observational biases.

Since all the FRBs used in \james\ were detected at $\sim 1.3$\,GHz, this model was perfectly appropriate. However, when incorporating lower-frequency observations --- i.e.\ the \icslow\ observations used here --- it becomes important to separate out the contribution of scattering $w_s$, which scales approximately as $w_s \sim \nu^{-4}$ \citep{Bhat2004,Day2020}, and can dominate the FRB width distribution at low frequencies.

The best measure of the scattering distribution of FRBs comes from the CHIME catalogue \citep{CHIME_catalog1_2021}. Using real-time injected bursts to estimate the effect of observational biases, these authors find that the true scattering time distribution at 600\,MHz, $\tau_{600}$, follows an approximate log-normal distribution with $\log \mu_s [{\rm ms}]=0.7$ and $\log \sigma_s=1.72$. We therefore use this result, and scale $\mu_s$ as 
\begin{eqnarray}
\log \mu_s (\nu) & = & 0.7 - 4 \left( \log \nu_{\rm obs} \,{\rm MHz} - \log 600\,{\rm MHz} \right). \label{eq:scat_scaling}
\end{eqnarray}
Thus our model for the total effective width, $w_{\rm eff}$, of FRBs becomes the quadrature sum of the intrinsic width $w_{\rm int}$, scattered width $w_{\rm scat}$, DM smearing width $w_{\rm DM}$, and sampling time $w_{\rm samp}$, i.e.\
\begin{eqnarray}
w_{\rm eff} & = & \sqrt{w_{\rm int}^2 + w_{\rm scat}^2 + w_{\rm DM}^2 + w_{\rm samp}^2}. \label{eq:width}
\end{eqnarray}

\begin{figure}
    \centering
    \includegraphics[width=\columnwidth]{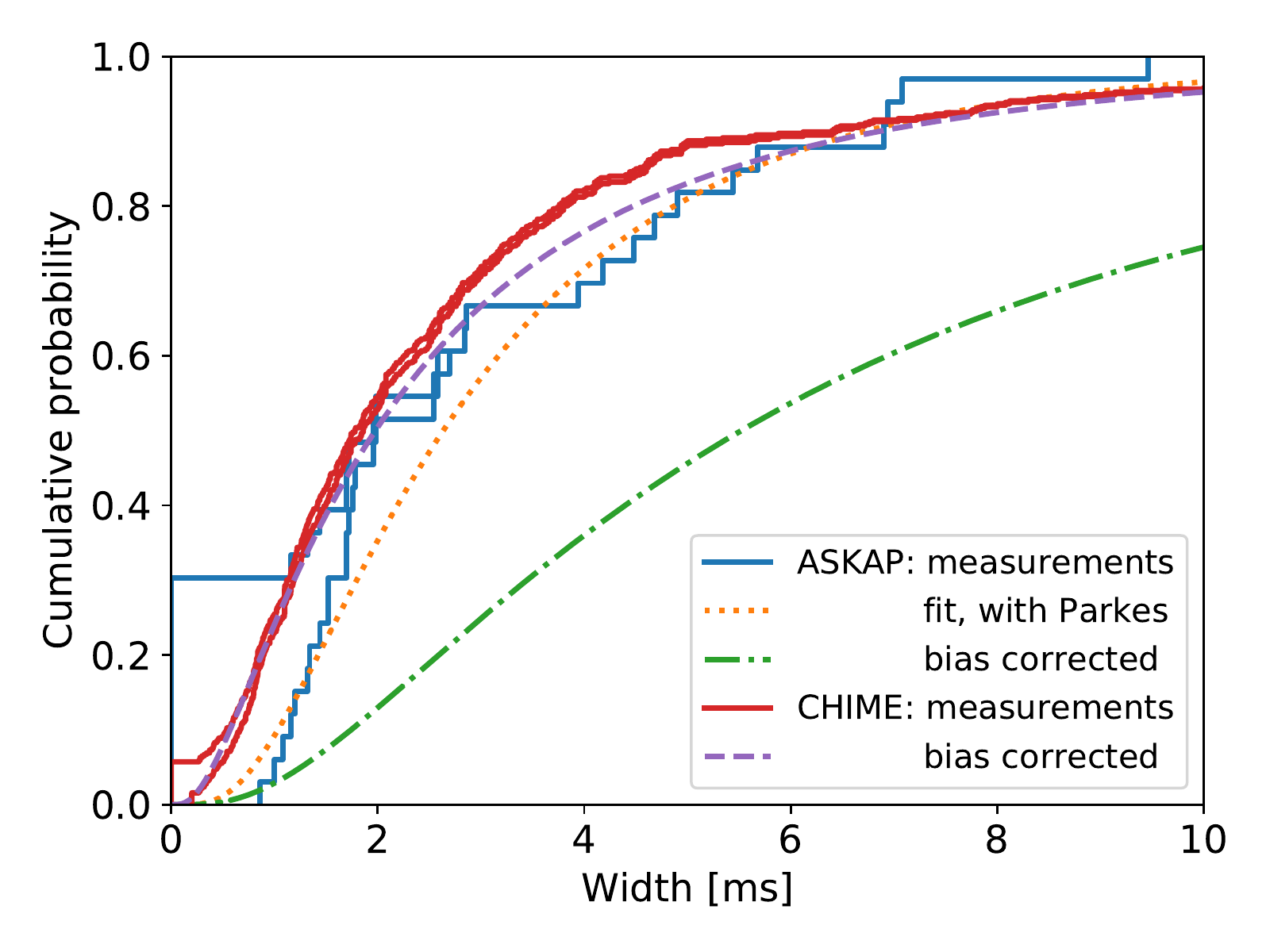}
    \caption{FRB width cumulative distributions. Shown are measurements from ASKAP \citep{Qiu2020} and CHIME \citep{CHIME_catalog1_2021}, with upper and lower lines calculated by assuming FRBs with upper limits have widths equal to zero and the limit value respectively. Also shown is a fit to data from ASKAP and Parkes \citep{Arcus2020}, and estimates of the bias-corrected (intrinsic) width distributions (\james). The ASKAP and Parkes data have not had the effects of scattering removed. Note that we use the total FRB widths, which are twice the reported Gaussian standard deviations.}
    \label{fig:intrinsic_widths}
\end{figure}

From \figref{fig:intrinsic_widths}, the measured and bias-corrected width distributions found by CHIME \citep{CHIME_catalog1_2021} are broadly consistent with the measurements of ASKAP and Parkes \citep[][\james]{Qiu2020,Arcus2020}, but narrower than the bias-corrected values \citep{James2022Meth}. This is an interesting result in-and-of itself, and assuming it is not due to some difference in the fitting methods, it may imply some frequency-dependent aspect of the FRB emission mechanism, or an unknown selection effect. It is also in contrast to the results of \citet{Gajjar2018}, who find that the intrinsic width of bursts from FRB~20211102 decreases with increasing frequency. For our purposes however, we simply retain the previous bias-corrected width distribution from \james, and add the contribution from scattering according to \eqref{eq:scat_scaling} and the parameters found by \citet{CHIME_catalog1_2021}.

\begin{figure}
 \centering
\includegraphics[width=3.5 in]{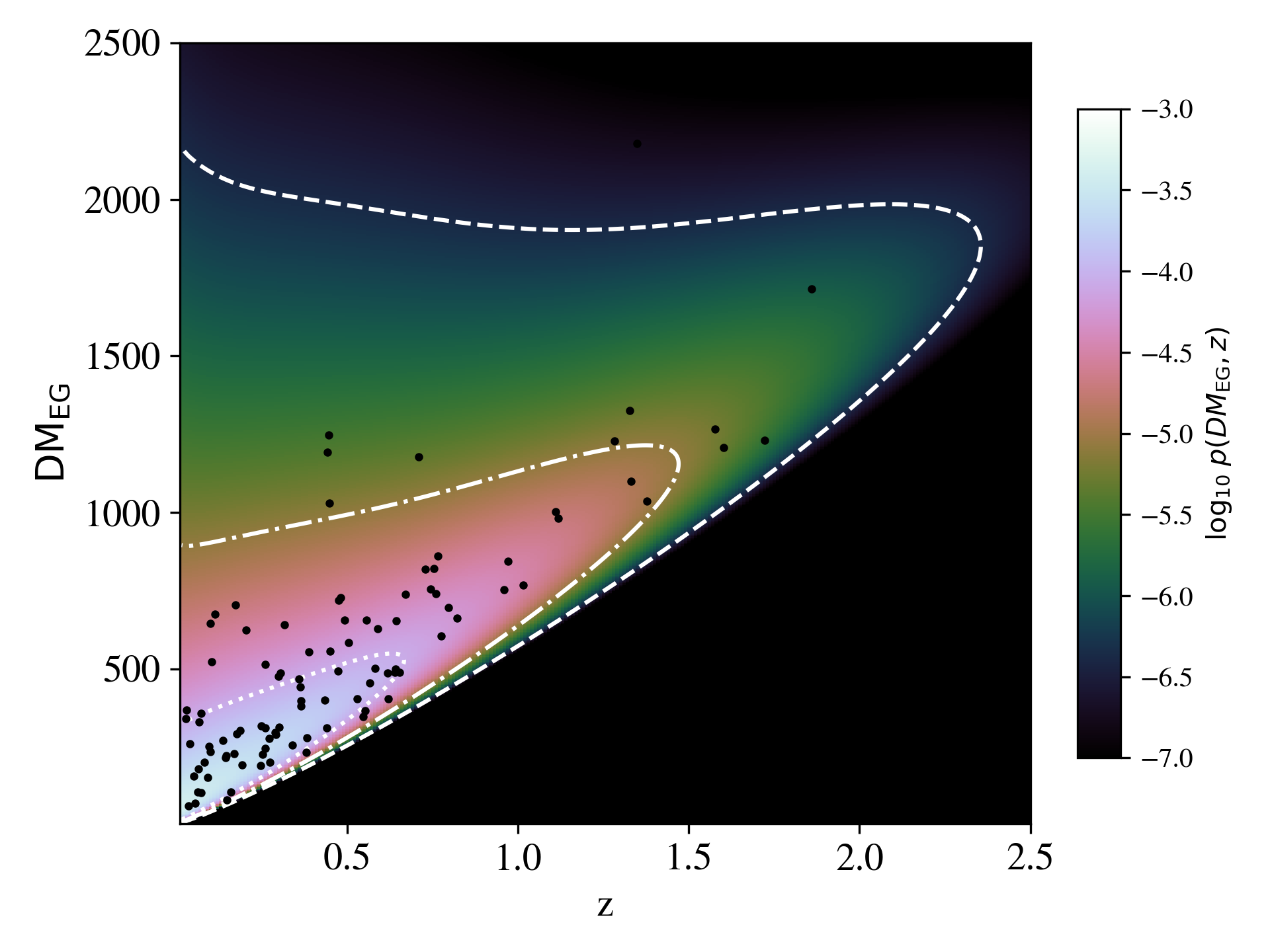}
\caption{The color image describes \pzdm\ 
for the forthcoming CRACO survey on the
ASKAP telescope for a fiducial set of 
model parameters (Table~\ref{tab:param}).
Overplotted are white contours enclosing
50\%\ (dotted),  90\%\ (dash-dot) and 99\%\ (dashed)
of the probability.  The black dots are a
Monte Carlo realization of the PDF for a 
random draw of 100 FRBs (Table~\ref{tab:MC}).
}
\label{fig:craco} 
\end{figure}

\begin{figure}
 \centering
\includegraphics[width=3.3 in]{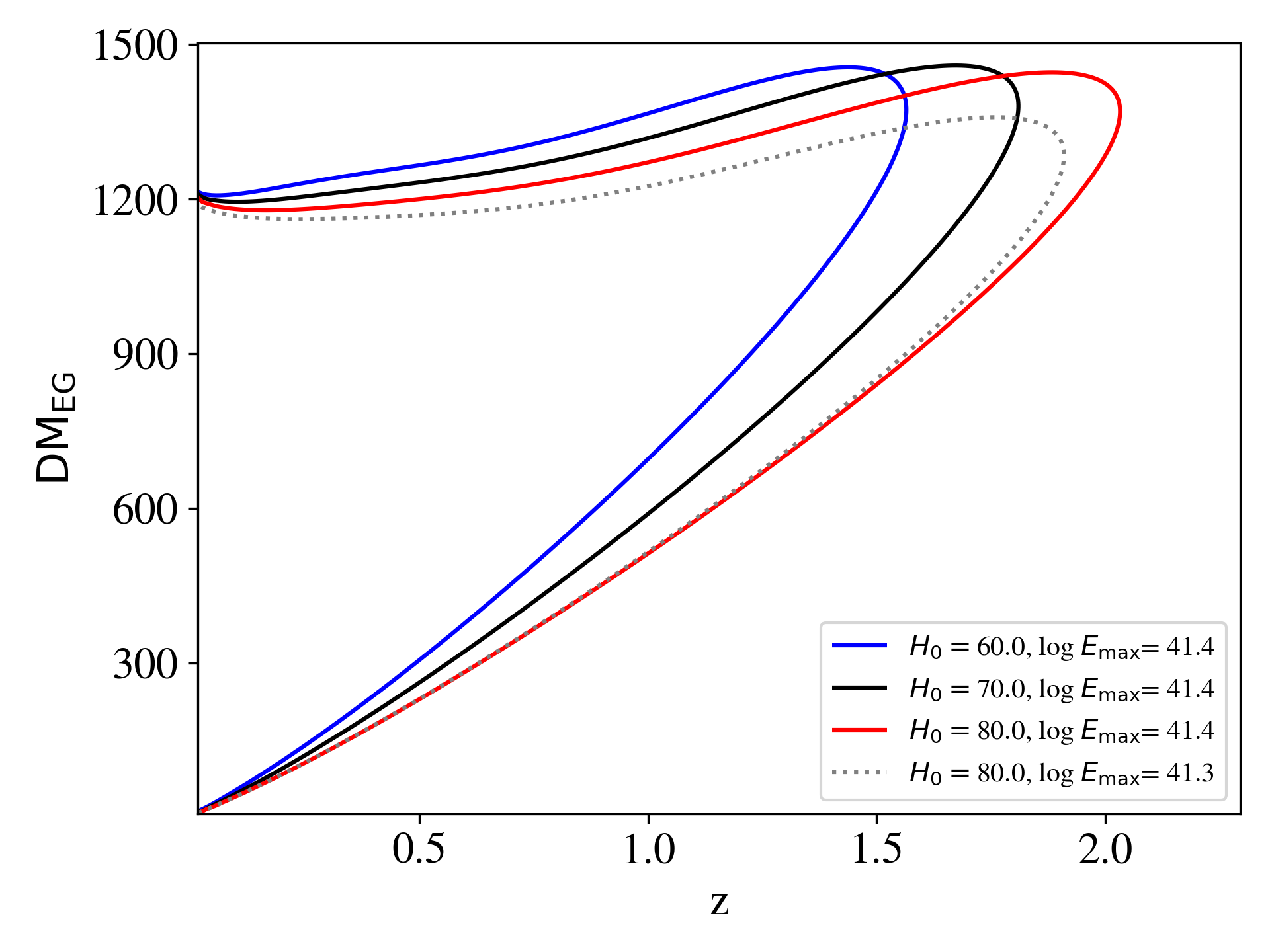}
\caption{The colored curves are the 95\%\
contours in the \pzdm\ space for the fiducial
CRACO model (Table~\ref{tab:param}) 
but with \Hnot\ varying from 60~\hubbunit\ (blue)
to 80~\hubbunit (red).
As \Hnot\ increases, the contours tilt towards
lower \dmeg\ values due to the \Hnot$^{-1}$
dependence of \dmacosmic\ (Equation~\ref{eqn:invH}).
They also extend to higher $z$ because
higher \Hnot\ implies a physically smaller universe,
i.e.\ one can observe an FRB with given energy
to higher $z$.
This effect, however, is partially degenerate
with the energetics of the FRB population
as described by the dotted line 
(a model with lower \emax).
}
\label{fig:vary_H0_zDM} 
\end{figure} 

\begin{figure}
 \centering
\includegraphics[width=3.3 in]{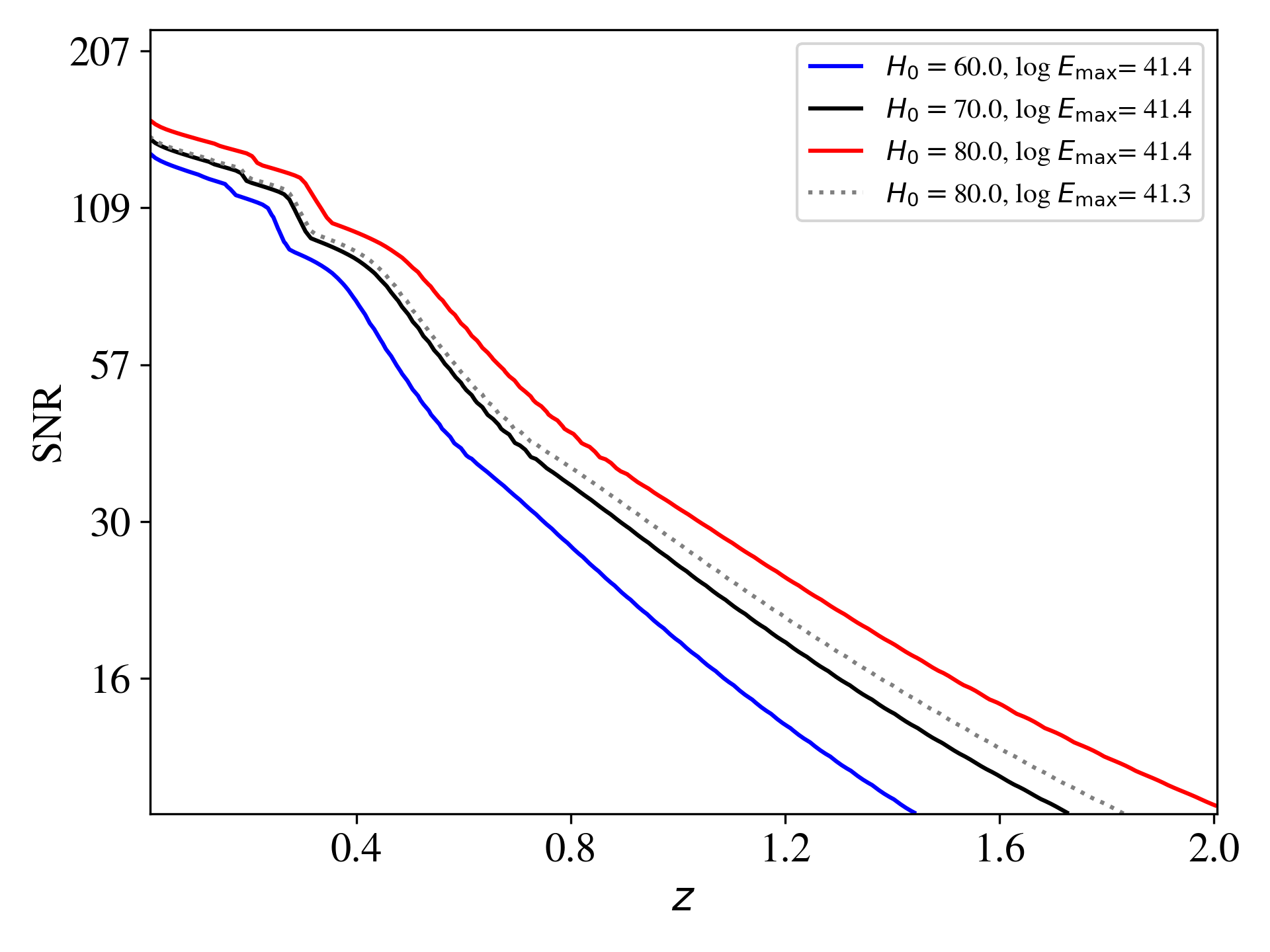}
\caption{Similar to Figure~\ref{fig:vary_H0_zDM}
except the curves are contours in the 
\pzs\ space (95\%\ of the events are expected
to occur below the lines).
While the models indicate significant 
\Hnot\ dependence, these are more nearly degenerate
with \emax\ than the results in the \pzdm\ space,
e.g, compare the solid, black curve with the dotted curve
which have significantly differing \Hnot\ and \emax.
}
\label{fig:vary_H0_sz} 
\end{figure}

\input{Tables/tab_model_params}

\section{Exploring the Model Dependencies on \Hnot}
\label{sec:explore}

In this section, we consider examples of the forward
model to gain intuition on the constraints 
for \Hnot\ imposed by
the observations as well as key
model degeneracies.
In the following, we assume properties for
future FRB surveys on the ASKAP telescope using the CRAFT COherent upgrade (CRACO) system.  Its characteristics follow the ICS (mid)
survey performed on ASKAP by the CRAFT project
but with approximately 4.4 times greater sensitivity due to the 
anticipated coherent addition of 24~antennas
(as opposed to the incoherent addition of
typically 25), and a slightly reduced bandwidth
(from 336\,MHz to 288\,MHz).

Figure~\ref{fig:craco} shows the \pzdm\ 
probability distribution function (PDF) for this 
CRACO survey and a fiducial set of model
parameters (Table~\ref{tab:param})
informed by \james. Overplotted 
is a Monte Carlo realization of 100 random FRBs
drawn from the 2D PDF.
These are, as expected, located primarily within the
90\%\ contour in PDF.
This Monte Carlo sample is analyzed in 
$\S$~\ref{sec:craco_mc_results} to perform a forecast
on the future sensitivity of FRB surveys to \Hnot.

The \pzdm\ PDF is highly asymmetric with a long
tail to large \dmeg\ values.
This asymmetry is driven by the predicted
tails in \dmcosmic\ due to the Poisson nature
of cosmic structure (e.g.\ large \dmcosmic\ values
from galaxy clusters)
and the adopted log-normal
PDF for \dmhost.
At the highest \dmeg\ values ($> 1500\, \dmunits$),
\pzdm\ tends towards {\it lower} redshift.
This counter-intuitive effect is due to the 
reduction in \snr\ by DM-smearing of the signal
combined with simple cosmological dimming
\citep[][\james]{Connor2019}.

Now we consider differences in \pzdm\ due to
variations in \Hnot.  
Figure~\ref{fig:vary_H0_zDM}
shows the 95\%\ contours in the \pzdm\ plane
for a range of \Hnot\ values and two choices
of \emax.  The results may at first 
seem counter-intuitive. In particular,
the models with higher \Hnot\ lean toward
{\it lower} \dmeg\ values in the \pzdm\ plane
even though $\dmcosmic \propto \omegabh$.
This occurs because we have held 
\omegabhh\ fixed (see $\S$~\ref{sec:DM})
such that increasing \Hnot\ {\it decreases}
\omegabh\ proportionally
and therefore \dmcosmic\ (Equation~\ref{eqn:invH})
and therefore \dmeg.
Because the tail to high \dmeg\ includes 
attributes of the host (\dmhost) 
and the distribution of baryons with the
Universe's cosmic web \citep[parameterized by $F$][]{Macquart2020},
the greatest constraining power 
on \Hnot\ is from the lower boundary of 
the contours.
This is evident in Figure~\ref{fig:craco}
where one notes the sharpness of \pzdm\ 
along the lower edge of the PDF contours.

Another important behaviour seen in 
Figure~\ref{fig:vary_H0_zDM} is 
that the contours `rotate' within the plane
as \Hnot\ varies.
All of the other model parameters that 
significantly affect \dmeg\ (e.g.\ those 
influencing \dmhost) tend to 
rigidly shift and/or widen the contours
parallel to \dmeg.
Therefore, there is significant constraining power
in the data for \Hnot\ without high
degeneracy.

The other notable effect of increasing \Hnot\
is that the contours extend to higher redshift.
With all other cosmological parameters fixed
(except $\Omega_b$),  a universe with
higher \Hnot\ is `smaller'.  Surveys with a 
given flux sensitivity can therefore observe
FRBs to higher redshift.
This secondary effect, however, is 
partially degenerate with the FRB luminosity function
and especially \emax.
Figure~\ref{fig:vary_H0_zDM} shows an additional
contour with $\Hnot=80 \, \hubbunit$
and an \emax\ value 20\%\ lower than the fiducial
value.  Lowering \emax\ reduces the
redshift extent of the 
$\Hnot = 80 \, \hubbunit$ to be similar to that
with a higher \emax\ and $\Hnot = 70 \, \hubbunit$,
although the contours remain offset. This also suggests some sensitivity to the functional form of the luminosity function of \eqref{eq:schechter}. We further investigate correlations between \Hnot\ and \emax\ when fitting to data in \secref{sec:results_correlations}.

This coupling of \Hnot\ and \emax\ manifests in the
other primary FRB observable: $s$.
Put another way, to the extent that energetics of
the FRB phenomenon are invariant with redshift
the analysis is effectively a standard candle.
We illustrate the model dependence in 
Figure~\ref{fig:vary_H0_sz}
which shows the PDF of \pzs\ for several choices
of \Hnot\ and \emax. 
There is a strong \Hnot\ dependence on the predicted 
distribution for $s$ as a function of redshift,
but the variance is nearly degenerate
with \emax, e.g. decreasing \Hnot\ by 10\,\hubbunit\ 
is nearly equivalent to lowering \emax\ by 0.1\,dex
(compare the black solid and dotted curves in 
Figure~\ref{fig:vary_H0_sz}).

\input{Tables/tab_ics_frbs}

\section{Observational Sample}
\label{sec:sample}

The FRBs analyzed here mainly draw from the same 
samples of \james\ and we refer the reader to that
manuscript for full details. Briefly, the three samples used are FRBs detected by the Murriyang (Parkes) Multibeam system \citep[\mb; e.g.][]{staveley1996parkesMultibeam,SUPERB1}, ASKAP when observing in Fly's Eye mode \citep[\fe;][]{Bannisteretal2017}, and ASKAP when observing in incoherent sum mode \citep[\ics;][ \icspaper]{Bannister2019}.
Here we describe updates to this data set, and the methods used to address bias in the data.

Our criteria aim to be inclusive in our data selection, in order to overcome the limitations from the small number of localized FRBs.
The studies presented in Appendix~\ref{sec:systematics} suggest that any systematic effects of doing so will be small compared to the statistical error due to small sample size. We expect to revise these criteria when more data become available. See \james{} for a discussion of observational biases against high-DM FRBs.

\subsection{New localized FRBs}

Since the publication of \james, the CRAFT survey has continued to observe commensally in incoherent sum mode. While observations are still ongoing, we include all FRBs detected up to Dec 31$^{\rm st}$ 2021. This adds 14 new FRBs to our sample. Their relevant properties are listed in Table~\ref{tab:ics_frbs}, while their detailed properties will be given in several works currently in preparation (\spirals; \icspaper, Gordon et al., in prep.).

\subsection{Addition of FRBs with higher \dmism}

\input{Tables/tab_fe_pks_frbs}

In \james, only FRBs with $\dmism<100$\,\dmunits\ were included in the analysis. This is because higher values of \dmism\ degrade sensitivity to FRBs, and it is too computationally expensive to calculate sensitivity for each individual FRB. Rather, the simulation uses the mean value of \dmism\ for the sample to calculate this observation bias, while using individual values of \dmism\ to calculate \dmeg\ for the purposes of likelihood evaluation.
This criterion previously rejected eight FRBs from \mb, and two FRBs from \fe. We show in Appendix~\ref{sec:ISM_effect} that this criterion can be relaxed somewhat, and we now include all previously excluded FRBs. This includes FRB~20010621, which we consider has sufficient excess DM beyond the estimated \dmism\ to be classified as an (extragalactic) FRB. These are listed in Table~\ref{tab:parkes_frbs}. In the future, when larger numbers of localised FRBs reduce statistical errors, the very small bias due to this approximation could become relevant, and this criterion may have to be revisited.

\subsection{Extension to other frequency ranges}

ASKAP/CRAFT observations in ICS mode are predominantly fully commensal. This means that FRBs may be detected in any of the four ASKAP observing bands, covering 600-1800\,MHz \citep{Hotan2021ASKAP}. Within each band, the precise choice of which 336 1\,MHz frequency channels are available to the CRAFT system also varies on a per-observation basis. Typically however, observations have clustered around two main frequency ranges near $900$\,MHz and $1.3$\,GHz, with a few further observations near 1.6\,GHz. We label these ranges \icslow, \icsmid, and \icshigh{} respectively. We therefore calculate \pn,
the probability of detecting a total of \Nfrb\ FRBs,
and \pszdm\ separately for each of these three frequency ranges, and treat these as independent surveys, with \Nfrb\ equal to
8 FRBs, 13 FRBs, and 1 FRB respectively. For computational simplicity, as per \james, within each survey we average the sensitivity over several observation-to-observation differences, such as the number of summed antennas (typically 25), observation frequency, beam configuration, system temperature, and time resolution, as well as \dmism{} as discussed above (however, \dmeg\ is calculated individually for each FRB).

The extension to frequencies beyond the nominal 1.3\,GHz addressed by \james\ also requires considering the effects of scattering separately from the intrinsic FRB width, as discussed in \secref{sec:scattering}. It does however allow the inclusion of FRB~20191001, which was excluded from \james{} as being the only FRB at the time of analysis to be discovered in \ics\ observations outside the 1.3\,GHz band.

\subsection{Consideration of host galaxy probability}
\label{sec:path}

The redshifts associated with each localised FRB are derived from observations of the host galaxy --- which necessarily requires a firm association of the FRB with that galaxy. The Probabilistic Association of Transients to their Hosts \citep[PATH;][]{path} gives a method to calculate posterior probabilities \pox\ of any given host galaxy association, while accounting for FRB localisation uncertainties. In the original analysis, seven of nine \ics\ FRB host galaxies associations were found with $\pox\ > 95$\%.

\citet{Bhandari+22} has performed an updated PATH analysis of three localized CRAFT FRBs, and reported a posterior probability \pox\ for the host association 
exceeding 90\%\ in each case.
In \icspaper\ we argue that one should
modify the standard PATH priors introduced by 
\citet{path} which increases the \pox\ values for these
and all previous FRBs from CRAFT/ICS. Thus all localised FRBs in our sample have posterior values of \pox\ of 90\% or greater, as listed in Table~\ref{tab:ics_frbs}.

\subsection{\ics{} FRBs with no hosts}
\label{sec:no_hosts}

The results and forecasts presented thus far
have implicitly assumed that we have observed
a complete and unbiased sample from the FRB surveys.
We recognize, however, that there is no perfect
FRB survey nor related follow-up efforts (e.g.\ to
obtain the FRB redshift). Of the FRBs included in Table~\ref{tab:ics_frbs}, six have no identified host. There are many potential reasons for this (numbers are for the current sample in \tabref{tab:ics_frbs}):
\begin{enumerate}
    \item the buffered data necessary for localisation was not available for technical reasons (3 FRBs); \label{list:buffer}
    \item the FRB host is obscured either by proximity to bright stars, or by high levels of dust extinction in the Milky Way (1 FRB); \label{list:obscured}
    \item the FRB host has not been observed yet, due to being too close to the Sun, or simply because the FRB is so recent that observations have not yet been completed (1 FRB); \label{list:unobserved}
    \item the FRB host cannot be identified amongst several candidate galaxies (no FRBs yet); \label{list:candidates}
    \item the FRB host is too distant or faint to be detected with ground-based follow-up observations (1 FRB). \label{list:distance}
\end{enumerate}
It is critical therefore that these effects do not introduce biases into our analysis.

Of the above, reasons \ref{list:buffer}, \ref{list:obscured} and \ref{list:unobserved} are clearly uncorrelated with the properties of the FRBs themselves, so that while missing these FRBs reduces our statistical power, using \pdm\ rather than \pzdm\ introduces no bias. Reason \ref{list:candidates} is a function of both the radio localization accuracy, and the number and properties of galaxies in the FRB field. While more distant FRBs are on-average dimmer \citep{Shannonetal2018}, and thus will have a greater statistical error on their localization, the correlation between SNR and $z$ is relatively weak; furthermore, the localization accuracy of \ics\ FRBs is typically dominated by systematics in FRB image alignment \citep{Day2021}, which are uncorrelated with FRB properties. However, since angular diameter distance is increasing over the redshift range of observed \ics\ FRBs, a constant angular resolution will result in a more-difficult host galaxy identification with increasing $z$, making it more likely to preferentially reject FRBs from high redshifts. Furthermore, one may not be able to obtain a sufficiently high-quality spectrum of the galaxy to confidently measure its redshift. Reason \ref{list:distance} is clearly correlated with redshift: an FRB follow-up observation probing to a limiting r-band magnitude of 22 might be insufficient to detect a 0.1\,$L^*$ galaxy beyond $z=1$ or a 0.01\,$L^*$ galaxy beyond a redshift of 0.1 \citep{Eftekhari2017}.

We deal with these biases here
by choosing a maximum extragalactic dispersion measure, DM$_{\rm EG}^{\rm max} = 1000$\,\pccc, beyond which detected FRBs are classified as unlocalised regardless of whether or not their host has been identified. To avoid bias in redshift, it is critical that this criterion is independent of $z$; however, these FRBs must be included in the calculation of \pdm\ to avoid bias in that parameter. 
Our localizations are sufficiently certain that we are not currently affected by reason \ref{list:candidates}. FRBs which are unlocalized for any other reason are also included in the calculation of \pdm\ --- rejecting these would not introduce a bias, but would reduce the statistical power of the sample. We show in Appendix~\ref{sec:unlocalised} how this procedure allows an unbiased measure of \Hnot. In total, six \ics\ FRBs are treated this way --- see \tabref{tab:ics_frbs}.

\subsection{Observation time, and low SNR bias}
\label{sec:snrbias}

The question of observational bias against low-SNR FRBs has a long history \citep{Macquart2018a,James2019a_source_counts}. An analysis of the measured SNR of \ics\ FRBs however reveals that the majority of FRBs with SNR$\lesssim 15$ have been undetected, resulting in a total FRB rate which is approximately half that expected (Shannon et al., in prep.). Under the simplifying assumption of a Euclidean slope of $(N_{\rm FRB} > \snr) \propto \snr^{-1.5}$, we expect that for every FRB detected with ${\rm SNR} \ge 15$, 1.15 are detected in the range $9 \le {\rm SNR} \le 15$, where ${\rm SNR}_{\rm th}=9$ is the nominal \ics\ detection threshold. However, \ics\ have detected 16 FRBs with ${\rm SNR} \ge 15$, and only 6 with ${\rm SNR} \le 15$, when 18.4 might be expected. Some of the missing low-SNR FRBs can be attributed to periods of high RFI ($\lesssim 10\%$ of the searches)  where the detection threshold had to be raised as high as ${\rm SNR}_{\rm th}=14$; however, in most cases this loss remains unexplained. We note that there is no evidence for such a bias in \fe\ data; nor does there appear to be a correlation between missing FRBs and properties such as DM or frequency, although our ability to probe this is affected by low sample numbers.

The assumption of a 50\% loss of FRBs due to a bias against low-SNR events is also backed up by calculations of the absolute \icsmid\ FRB rate (Shannon et al., in prep), which was not available in time for use in \james. The expected rate of \icsmid\ FRBs is relatively model-independent, since the frequency range is almost identical to --- and the sensitivity lies between --- \fe\ and \mb\ observations. This reveals a detection rate which is approximately half that expected, which is consistent with the low-SNR bias described above.

We account for this issue therefore by taking the absolute observation time of \ics\ observations measured by Shannon et al.\ (in prep), and divide by half, to represent the 50\% loss of detection efficiency. Importantly, the measured FRB rate of \icslow\ and \icshigh\ observations allows us to have better statistical inference on the frequency-dependent rate parameter $\alpha$, which aside from a prior based on the results of \citet{Macquart2019a}, was largely unconstrained in \james. In Appendix~\ref{sec:snr_problem}, we demonstrate that missing FRBs in this small SNR range does not cause any significant bias in the determination of \Hnot, so we retain the threshold of ${\rm SNR}_{\rm th}=9.0$.

\section{Results}
\label{sec:results}

\begin{figure}
    \centering
    \includegraphics[width=3.3in]{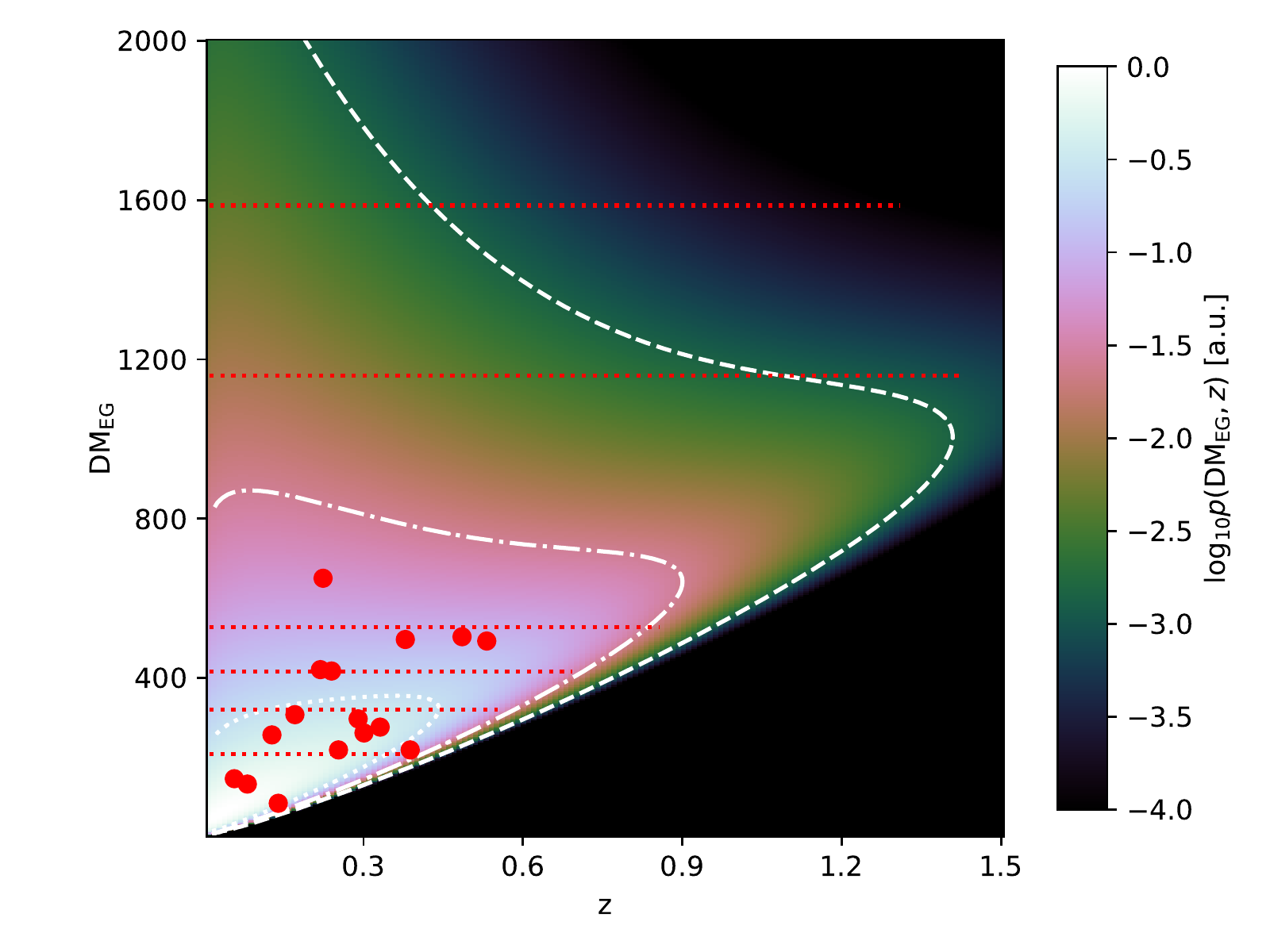}
    \caption{The z--DM distribution of FRBs (shading) using best-fitting model parameters, summed over the \icslow, \icsmid, and \icshigh\ samples. Also shown are FRBs with (red solid circles) and without (red dashed lines) host galaxy redshifts. 
    The latter are drawn at their estimated values of \dmeg\ out to $z_{99}$, i.e.\ encompassing 99\%  of their likelihood in \pzgdm\ for their \dmeg\ values. 
    White contours of dotted, dash-dot, and dashed lines
    encompass 50\%, 90\%, and 99\% of the probability density of \pzdm\ respectively.}
    \label{fig:bestfit_zdm}
\end{figure}

\input{Tables/tab_cube_params}

We evaluate \pn\ for each survey, and \pszdm\ for each FRB in that survey, over a seven-dimensional cube of parameters, with values given in \tabref{tab:cube_params}. Posterior probabilities are calculated from the resulting product over all FRBs and surveys using uniform priors over the simulated parameter ranges, while confidence intervals on those parameters are constructed using the prescription of \citet{FeldmanCousins1998}. Our results are given below. The best-fitting z--DM distribution for ASKAP FRBs is shown in \figref{fig:bestfit_zdm}.

\subsection{\Hnot}
\label{sec:resultsH0}

\begin{figure}
    \centering
    \includegraphics[width=3.3in]{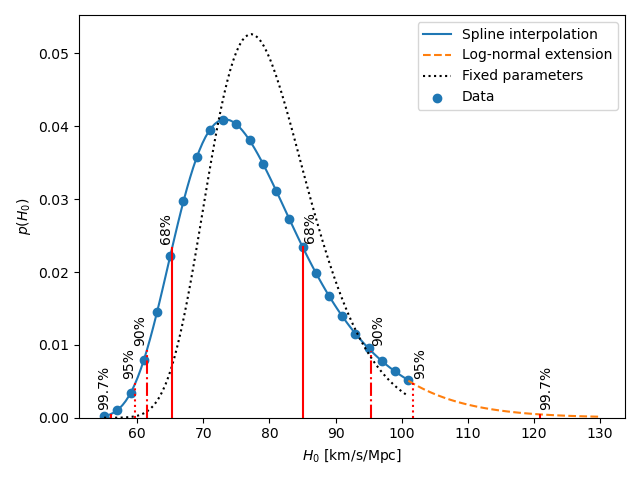}
    \caption{Posterior probability on \Hnot, using cubic splines (blue solid curve) fitted to data points (circles), using uniform priors over the simulated range, and extrapolated to higher values of \Hnot\ using a log-normal fit (orange dashed curve). Confidence intervals are also shown (vertical red lines). The posterior probability when fixing all other parameters to their best-fit values is also shown (grey dotted curve).}
    \label{fig:1DH0}
\end{figure}

Our posterior probability distribution for \Hnot\ is given in \figref{fig:1DH0}. We simulate only up to \Hnot=103\,\hubbunit: in the range above 80\,\hubbunit, we find the probability to be decreasing as a log-normal to better than 1\% relative accuracy, so we save significant compute time and extrapolate results to \Hnot=130\,\hubbunit. Our best-fit value is \Hnot=\result\,\hubbunit. While this agrees with values of \Hnot\ derived from near-Universe measures, it is also consistent within $1\,\sigma$ of indirect values derived from e.g.\ the CMB \citep{SNOWMASScosmology}.

Our constraint on \Hnot\ is not symmetric --- we derive a relatively sharp lower boundary, with looser constraints on large values of \Hnot. This is the result of what we term `\effect', whereby large excess DMs above the mean can be induced by intersection with galaxy halos or host galaxy contributions, but even voids contribute a minimum \dmcosmic. 
Thus, probability distribution \pzgdm\ has a sharp lower cutoff, or `cliff'.
Low values of \Hnot\ thus imply a higher minimum DM as a function of $z$, and when this minimum is contradicted by even a single measured FRB, those values of \Hnot\ can be excluded. Higher values of \Hnot\ however do not suffer such a large penalty due to the long tail of the \pdmgz\ distribution.

\figref{fig:1DH0} also shows the constraint on \Hnot\ we would derive if all other FRB parameters are fixed to their best-fit values. This demonstrates the importance of performing a multi-parameter fit and marginalising over nuisance parameters. Even in the case that the fixed values of FRB parameters are well-guessed (i.e.\ at the best-fitting values used here), ignoring the confounding effects of uncertainties in these parameters nonetheless leads to a biased and artificially too-precise estimate of \Hnot.

\begin{figure}
    \centering
    \includegraphics[width=3.3in]{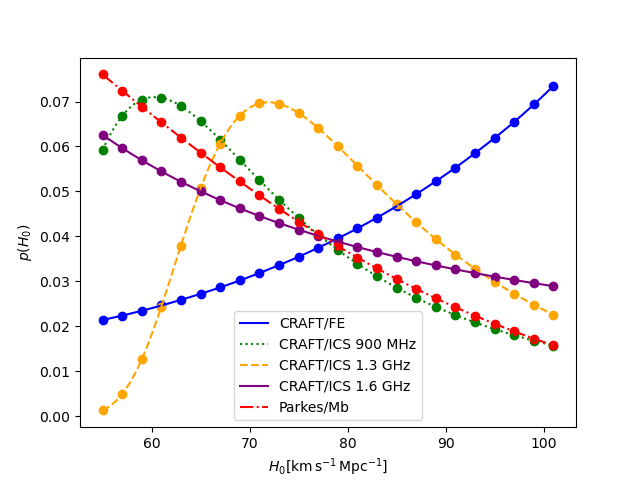}
    \caption{Posterior probability on \Hnot, using cubic splines (solid curves) fitted to data points (squares), using uniform priors over the simulated range, for each of the five FRB surveys used in this work.}
    \label{fig:H0BySurvey}
\end{figure}

It is also interesting to examine the constraints on \Hnot\ from each FRB survey individually. This is shown in \figref{fig:H0BySurvey}. The two surveys with large numbers of localised FRBs --- \icslow\ and \icsmid\ --- provide the dominant constraints, as expected. That \icsmid\ is better fit by a higher value of \Hnot\ is due to \effect\ and FRB~20190102, which has a very low DM of 322.2 \dmunits\ for its redshift of 0.378. That the single localised FRB of \icshigh{} provides a comparable amount of information to \fe\ and \mb, with 26 and 28 FRBs each, illustrates the importance of localised FRB samples when constraining \Hnot.

\subsection{Constraints on other parameters}
\label{sec:results_others}

\begin{figure*}
    \centering
    \includegraphics[width=3.3in]{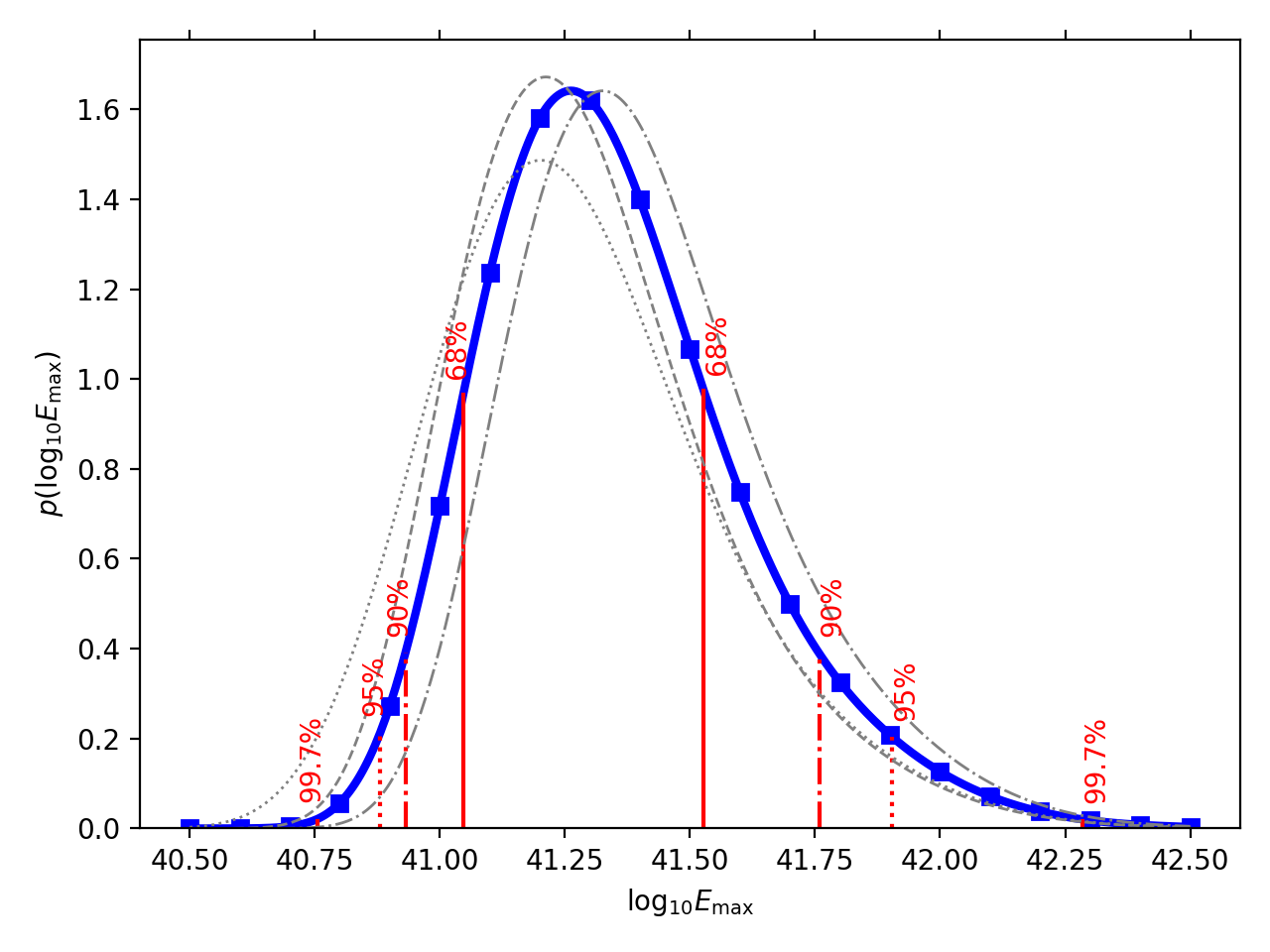}
    \includegraphics[width=3.3in]{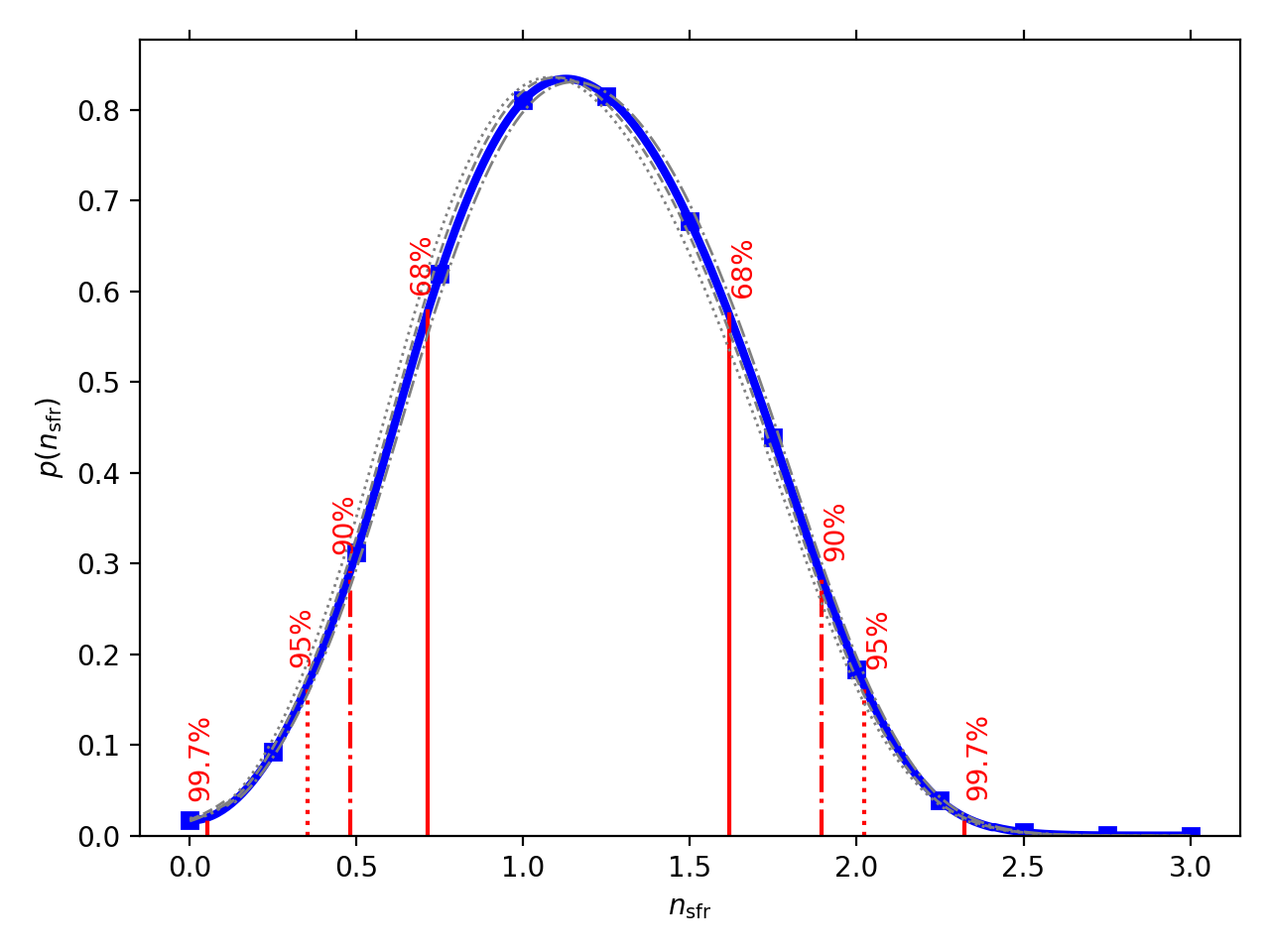} \\
    \includegraphics[width=3.3in]{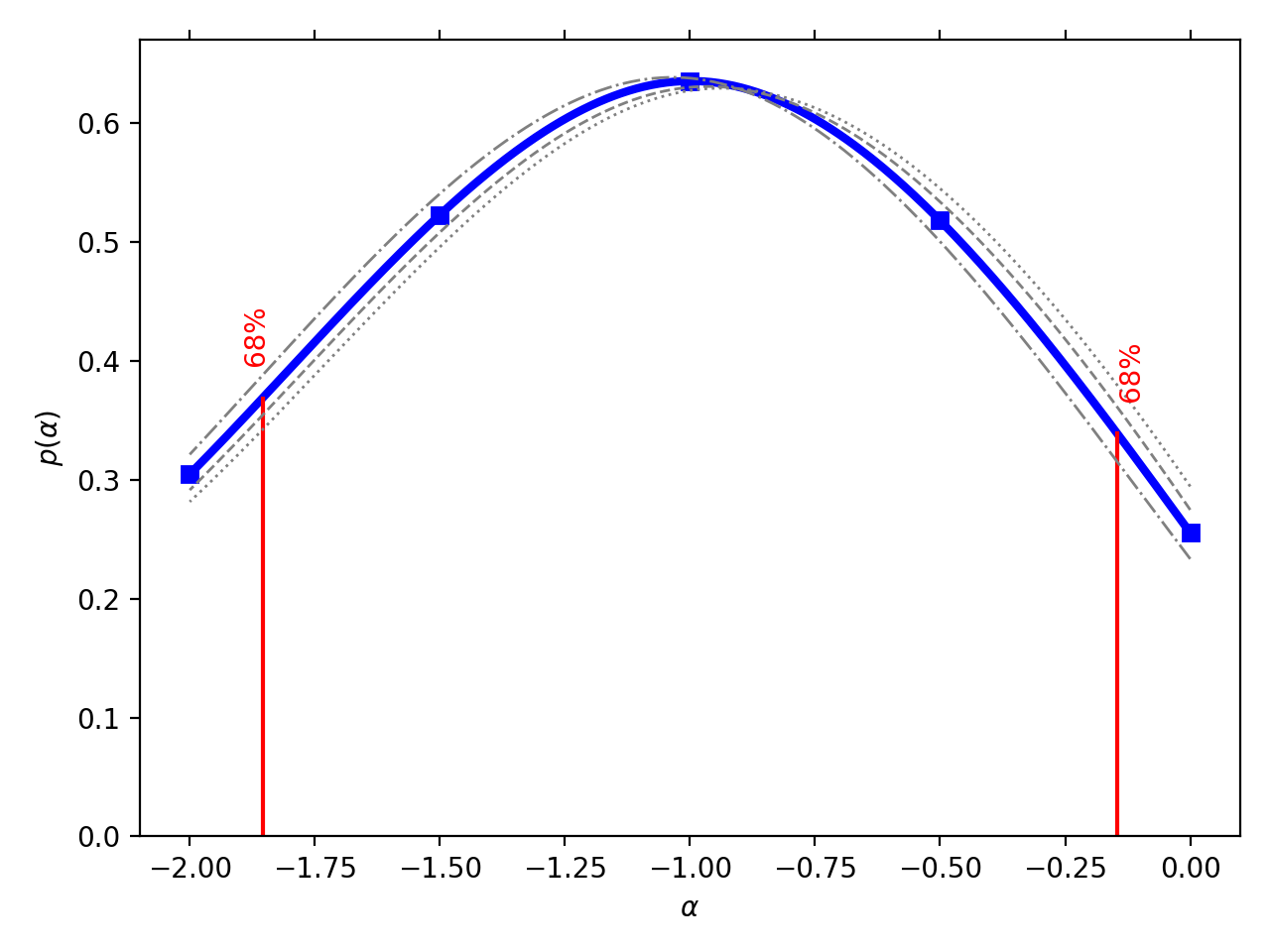}
    \includegraphics[width=3.3in]{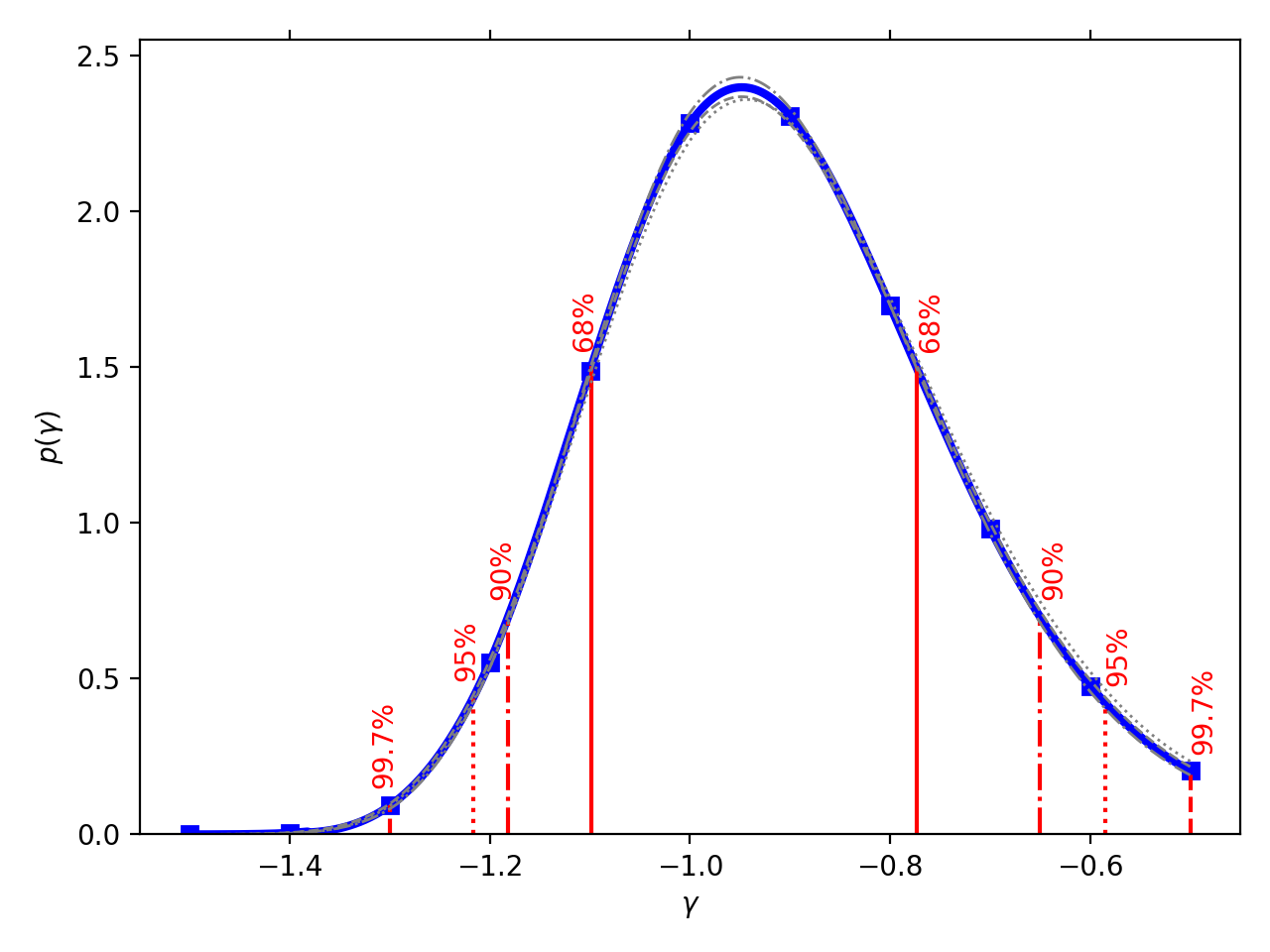} \\
    \includegraphics[width=3.3in]{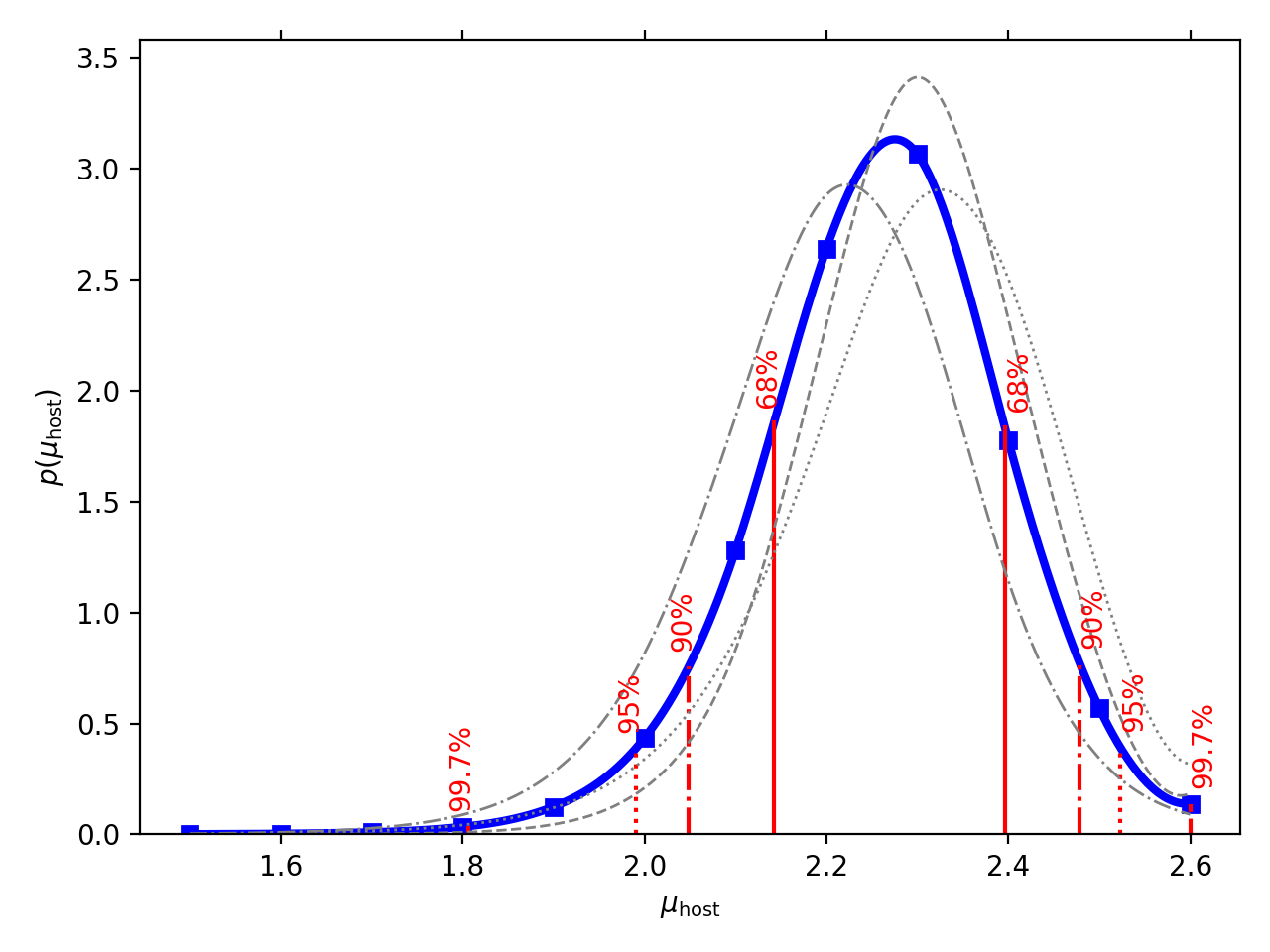}
    \includegraphics[width=3.3in]{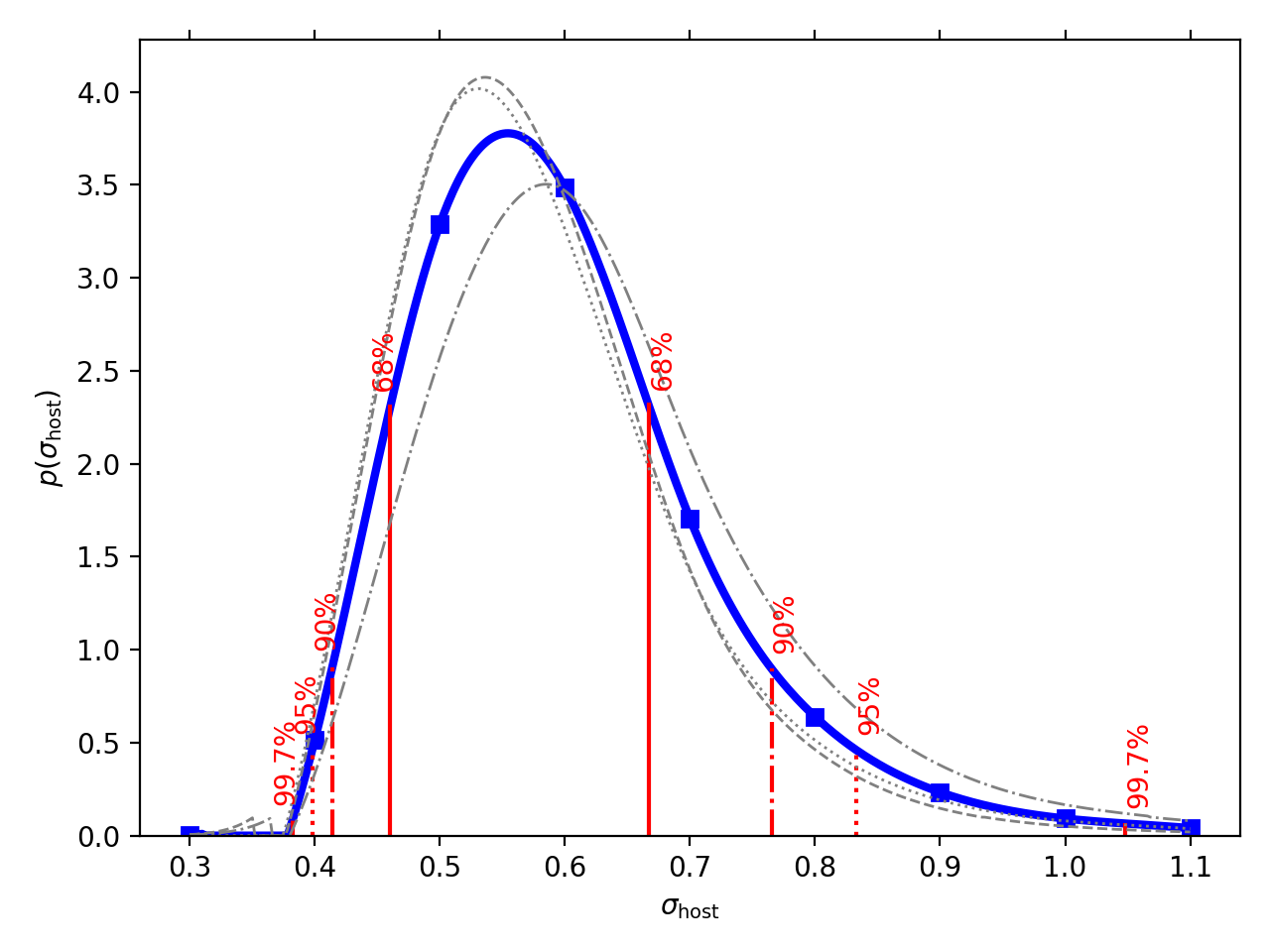}
    \caption{Posterior probabilities on the six other estimated parameters, using a prior for \Hnot\ based on CMB and SN1a results (blue curves; see text). 
    Also shown are posterior probabilities with no prior on \Hnot\ (grey dotted curves) and with \Hnot\ fixed to 67\,\hubbunit (grey dashed curves) and 73\,\hubbunit (grey dash-dot curves).}
    \label{fig:1Dother}
\end{figure*}

\input{Tables/tab_extended_results}

Besides constraints on \Hnot, our addition of new localized FRBs yields greater statistical power to constrain the other fitted parameters --- $\log_{10} E_{\rm max}$, $\alpha$, $\gamma$, $n_{\rm sfr}$, $\mu_{\rm host}$, and $\sigma_{\rm host}$ --- while also accounting for the confounding effect of allowing \Hnot\ to vary. However, since our constraint on \Hnot\ using FRB data only is significantly less than that of other measurements, we apply a prior on \Hnot\ which is flat between the best-fit CMB and SN1A values (67.4 and 73.04 \hubbunit), and falls away as a Gaussian on the lower/upper regions with the respective uncertainties of those measurements (0.5 and 1.42 \hubbunit).

Posterior probability distributions on these parameters are shown in \figref{fig:1Dother}, while
confidence limits on all parameters are reported in \tabref{tab:confidence_intervals}. For comparison, we also show results with no prior on \Hnot, which give worse constraints; and also results when assuming \Hnot\ is fixed to either the value of 67.4 obtained by \citet{PlanckCosmology2018} or 73.04 from \citet{Riess2021}, representing the improvement in accuracy should \Hnot\ be known exactly.

We confirm the result of \citet{James2022Lett} that the FRB population exhibits cosmological source evolution consistent with the star-formation rate, excluding no source evolution ($n=0$) at $3\,\sigma$. 
This is in-line with the expectations from models predicting a close association between star-forming activity and FRB progenitors, in particular young magnetar models \citep[e.g.][]{Metzger2017magnetar}, but does not exclude that a fraction of FRBs could arise from channels with a significant ($\sim$Gyr) characteristic delay from star formation, such as mergers. It does exclude that the majority of FRB progenitors have a cosmologically significant delay with respect to star formation. We observe that other results in the literature that analyse FRB population evolution \citep[e.g.][]{Cao2018delayedmergers,Locatelli2019VVmax,Arcus2020,Bhattacharyya2022} tend to assume a 1--1 z--DM (i.e.\ a purely linear) relationship, and/or fix values of other FRB population parameters, which are assumptions which we do not make.

We also find an increased value of $\muhost=2.27_{-0.13}^{+0.12}$, i.e.\ a median host DM of $186^{+59}_{-48}$\,\dmunits (the corresponding mean DM is 240\,\dmunits), which is significantly greater than the usually assumed value of 100\,\dmunits\ found in the literature. This may not reflect entirely upon the actual FRB host galaxy: our fit to \muhost\ will include any error in our assumed mean value of \dmhalo=50\dmunits, and some component of \sigmahost\ will include scatter about that mean, and also errors in \dmism. However, since our used values of \dmhalo=50\dmunits, and NE2001 for \dmism, are typical of the literature, it does suggest that the average work on FRBs is underestimating some combination of \dmhost, \dmism, and/or \dmhalo, and thus over-estimating \dmcosmic. Other results, such as the excess DM of $\sim900$\,\dmunits\ observed for FRB20190520B by \citet{Niu2022}, and the suggestion of correlation between the locations of CHIME FRBs and large-scale structure at an excess DM of $\sim400$\,\dmunits \citep{CHIME_2021_structure}, support this conclusion.

The slope of the intrinsic luminosity function, $\gamma$, is found to be $-0.95^{+0.18}_{-0.15}$ --- consistent with the observed high-energy slope of the luminosity functions of known repeating FRBs, e.g.\ $\gamma = -0.85 \pm 0.3$  \citep{Li2021_FAST_121102}, $-0.88 > \gamma > -1.29$ \citep{Jahns2022_Arecibo121102}, and $\gamma = -1.04\pm0.02$ \citep{Hewitt2021} for FRB~20121102A. This is consistent with, though not sufficient proof of, apparently once-off FRBs being simply the high-energy tails of intrinsically repeating objects.

The posterior distribution of $\alpha$ drops only to approximately half its peak value over our simulated range ($-2 \le \alpha \le 0$), suggesting an uncertainty of $\pm 0.85$. Thus we should consider that we have used a uniform top-hat prior on $\alpha$. Nonetheless, unlike \james, we have significant discrimination power on $\alpha$. Our best-fit value of $\alpha=-1.0 \pm 0.85$ is consistent with the result of \citet{Macquart2019a}, who find $\alpha=-1.5_{-0.3}^{+0.2}$ under the assumption that each FRB is characteristically broadband with a spectral slope, which (as argued in \james) should be revised to $\alpha=-0.65^{+0.2}_{-0.3}$ under the `rate approximation' used here. This disfavours the results of \citet{CHIME_catalog1_2021} and \citet{Farah2019} --- which do not include the effects of observational bias --- that there is no increase of the FRB rate at decreasing frequency.

We conclude this section by noting that none of the above results are strongly dependent on our choice of prior on \Hnot\ (none; based on existing literature; or fixed), with the greatest effect being on \emax\ and \muhost.

\subsection{Correlations with other parameters}
\label{sec:results_correlations}

\begin{figure*}
    \centering
    \includegraphics[width=3.3in]{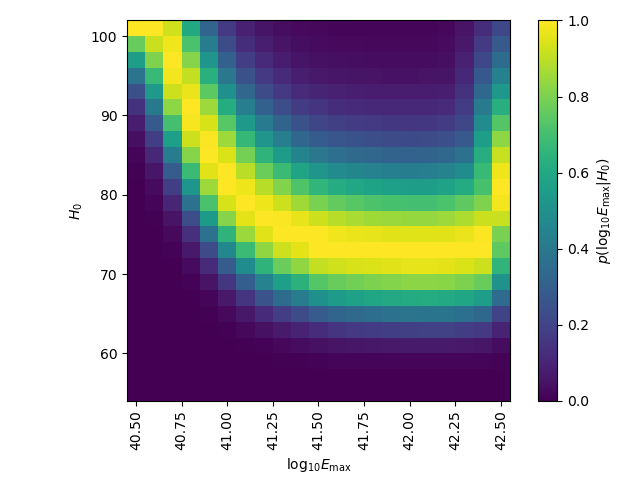}
    \includegraphics[width=3.3in]{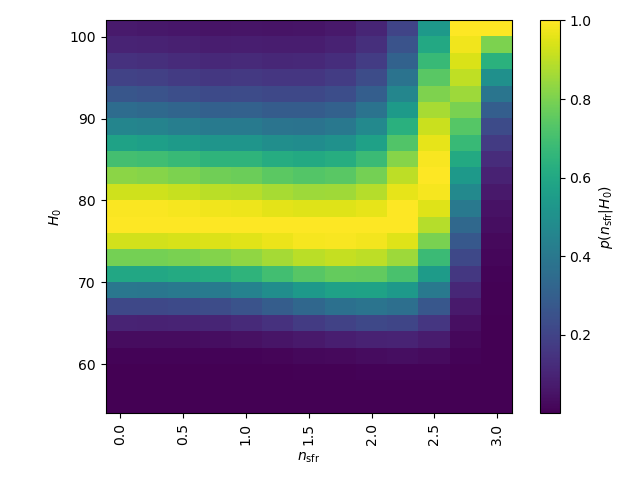} \\
    \includegraphics[width=3.3in]{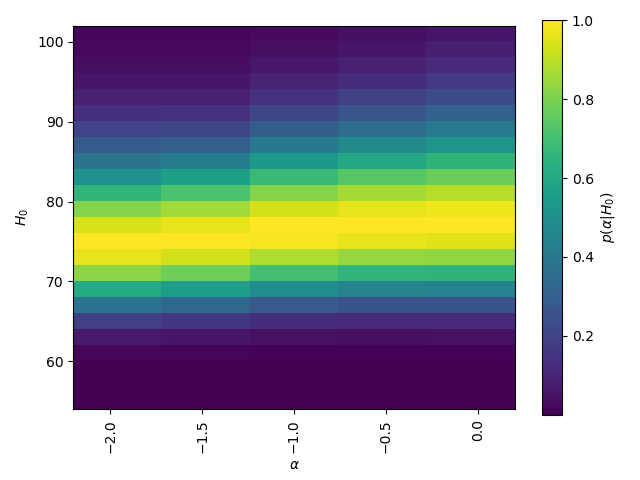}
    \includegraphics[width=3.3in]{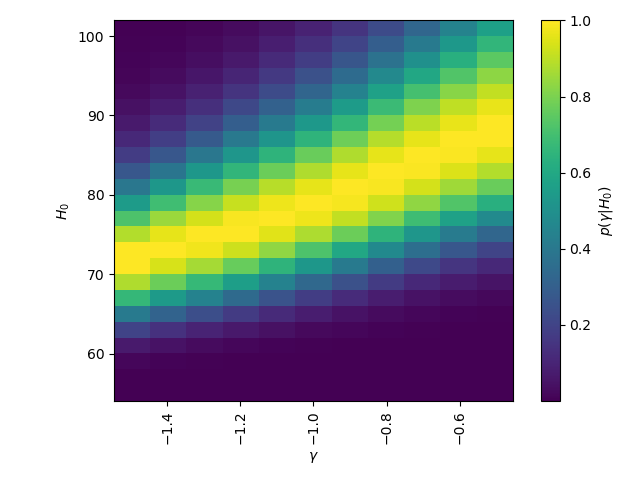} \\
    \includegraphics[width=3.3in]{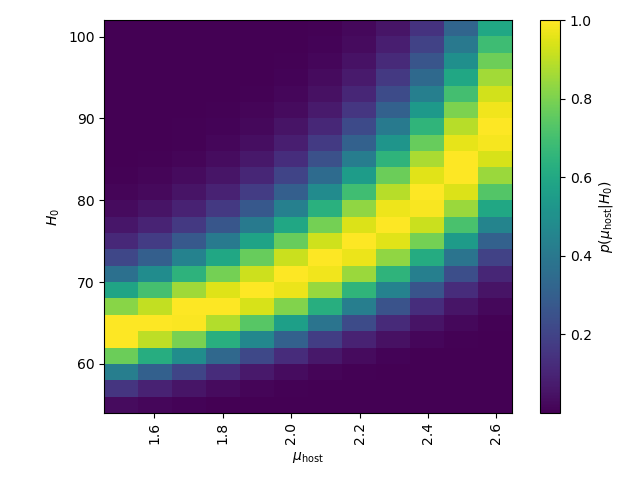}
    \includegraphics[width=3.3in]{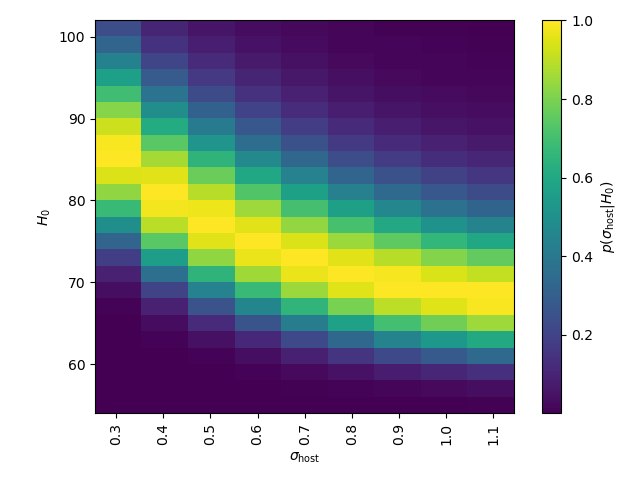}
    \caption{Conditional posterior probabilities on \Hnot\ when allowing only single parameters to vary from their best-fit values.
    These plots describe degeneracies in the results between 
    \Hnot\ and the other model parameters.}
    \label{fig:2DCorrelations}
\end{figure*}

We illustrate the correlations between parameters in \figref{fig:2DCorrelations}. We only show results for correlations between \Hnot\ and other parameters, since \james\ has already analysed other correlations. For each plot, we hold all other parameters constant at their best-fit values and plot the conditional probability $p(H_0|X)$ for each parameter $X$. 
From \figref{fig:2DCorrelations}, \Hnot\ is correlated with all modelled parameters, emphasising the importance of jointly fitting them.

Some correlations are readily understood. As \muhost\ increases, the implied \dmcosmic\ decreases, favouring smaller distances being travelled by the FRB --- and hence larger values of \Hnot. The strong negative correlation of \Hnot\ with low values of \emax\ is because smaller values of \emax\ require that distances to localised FRBs decrease to allow these FRBs to be detectable, which requires larger values of \Hnot. However, once \emax\ is sufficiently large (\lemax$>41.5$), there is essentially no correlation. The sharp increase of \Hnot\ for \lemax$> 42.5$ --- and for \sfrn\ $> 2.5$ --- is driven by the \mb\ sample, where for these extreme values of \lemax\ and \sfrn, a large \Hnot\ is required to reduce the DM of the otherwise very large number of distant FRBs that \mb\ would be able to detect.

As $\gamma$ increases, more FRBs are generated near \emax, and are thus visible from larger distances (higher $z$); while the effect of the experimental bias against high DMs is reduced (higher DM for a given $z$). It turns out for this data set that the latter effect is more important, and a positive correlation arises since increasing \Hnot\ reduces the expected DM.

That \sigmahost\ is anti-correlated with \Hnot\ –-- particularly for low values –-- can be understood via \effect. Reducing \sigmahost\ narrows the distribution of \dmeg\ about the Macquart relation. The reduced extent of the high DM tail still allows for excess-DM FRBs above the relation, albeit with reduced probability; however, a small reduction in \sigmahost\ massively decreases the likelihood of observing FRBs below the Macquart relation. To model this requires increasing \Hnot, since that pulls the Macquart relation downward, as shown in \figref{fig:vary_H0_zDM}.

For most values of \sfrn\ there is a very slight anti-correlation with \Hnot, since increasing both parameters can act to increase the fraction of high-redshift FRBs. However, for $\sfrn \gtrsim 2.5$, there is a strong positive correlation. This is largely driven by \pn, where increasing \Hnot\ decreases the volume in which very large numbers of FRBs would otherwise be predicted.

The slight positive correlation between the spectral rate parameter $\alpha$ and \Hnot\ is a combination of multiple minor effects, the sum of which has little impact on the determination of \Hnot. The influence of $\alpha$ will be constrained in the future by including FRB data from a wider range of frequencies.


\section{Forecasts --- CRACO Monte Carlo}
\label{sec:craco_mc_results}

\begin{figure}
    \centering
    \includegraphics[width=3.3in]{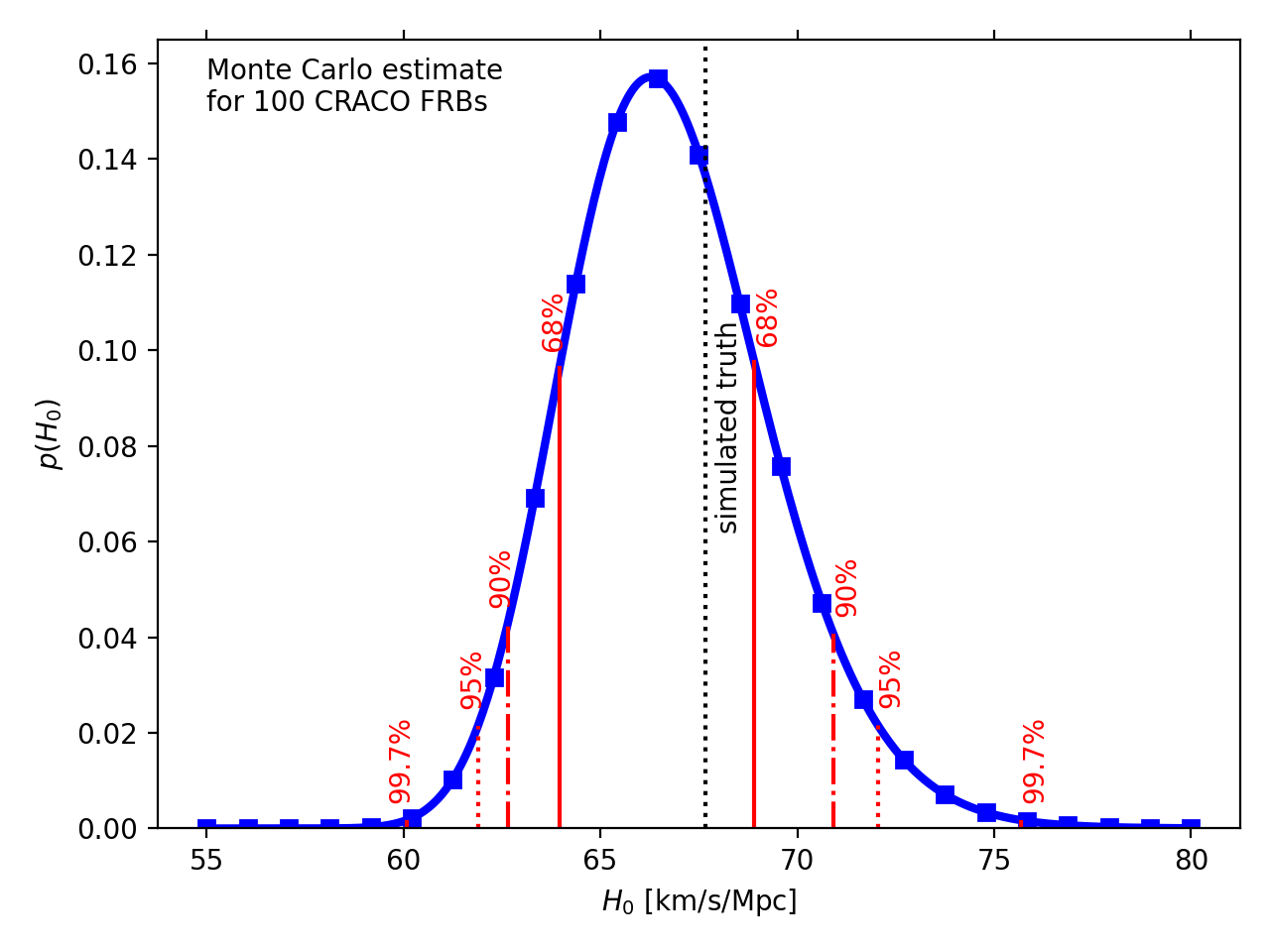}
    \caption{Posterior distribution on \Hnot\ calculated from our Monte Carlo sample of 100 simulated FRBs that would be detected by the ASKAP CRACO upgrade (blue). Also shown are confidence intervals (red; labelled), and the simulated true value (black vertical line). Results with a uniform and Gaussian prior on $\alpha$ are indistinguishable.} 
    \label{fig:craco_results}
\end{figure}

Our limit on \Hnot\ using the 16 localised and 60 unlocalised FRBs is not sufficiently constraining to discriminate between direct measurements from the local Universe and indirect measurements from the early Universe. However, in the near future, several experiments promise to greatly increase the number of localised FRBs. In particular, CRACO aims to implement a fully coherent image-plane FRB search during 2022.\footnote{https://dataportal.arc.gov.au/NCGP/Web/Grant/Grant/LE210100107}. The searches will be undertaken commensally with all other ASKAP observations.

We model CRACO by assuming $N_{\rm ant}=24$ of ASKAP's 36 antennas are coherently added over a $\Delta \nu=288$\,MHz bandwidth. Scaling the 22\,Jy\,ms detection threshold to a 1\,ms burst of the $\Delta \nu=336$\,MHz Fly's Eye survey \citep{James2019cSensitivity} by $N_{\rm ant}^{-1} (\Delta \nu)^{-0.5}$ gives an estimated detection threshold of $F_{\rm th}=0.99$\,Jy\,ms. Assuming a Euclidean dependence of the FRB rate on sensitivity (i.e.\ $R \propto F_{\rm th}^{1.5}$), the Fly's Eye rate of 20 FRBs in 1427 antenna days of observing predicts a 100-fold rate increase to 1.5 FRBs/day at high Galactic latitudes. This is then reduced by telescope down-time, RFI, time spent observing in the Galactic plane, and other efficiency losses in both radio observing and optical follow-up observations to identify host galaxies.
Here, we use 100\,FRBs to nominally represent the first year's worth of CRACO observations. These FRBs are listed in Table~\ref{tab:MC}, since they may be useful for other CRACO-related predictions, e.g.\ for gauging the requirements of optical follow-up observations.

We draw parameters $s$, DM, and $z$ randomly from the simulated distribution of \pszdm\, assuming Monte Carlo truth values of $\Hnot=67.66$\,\hubbunit and best-fit FRB population parameters from \citet{James2022Lett}. These FRBs are plotted in \figref{fig:craco} and listed in \tabref{tab:MC}. We repeat our calculation of \pszdm\ over the multidimensional grid of parameters given in \tabref{tab:cube_params}, albeit limiting this to a more constrained range of \Hnot. The resulting Bayesian posterior probability distribution on \Hnot\ is given in \figref{fig:craco_results}.

For this particular FRB sample, we find $\Hnot=66.28_{-2.3}^{+2.6}$\,\hubbunit, consistent with the MC truth value of 67.66\hubbunit. Importantly, the statistical uncertainty of $\sim2.45$\hubbunit --- which includes the increased variance when fitting for FRB population parameters --- would provide 2.5\,$\sigma$ evidence to discriminate between the $\sim6$\, \hubbunit\ difference in estimates of \Hnot\ \citep{SNOWMASScosmology}.
Assuming only statistical errors, to achieve $5\,\sigma$ discriminatory power would require 400 localised FRBs.

This suggests that near-future FRB observations --- and perhaps only a single year's worth of 100\% efficient observations with the CRACO upgrade on ASKAP ---
will be able to help resolve the current discrepancy in the two measurements of \Hnot. It also motivates a careful treatment of potential systematic errors, both in the FRB sample used, and in the cosmological and FRB population model. We discuss such errors in \secref{sec:uncertainties}.

\input{Tables/tab_MC}

\section{Discussion}
\label{sec:discussion}

\subsection{Comparison to other estimates of \Hnot\ with FRBs}
\label{sec:comparison}

\newcommand{\Wu}{Wu22}
\newcommand{\HS}{HS22}

Both \citet{WuH02022} (\Wu) and \citet{HagstotzH02022} (\HS) use measurements of FRBs to constrain \Hnot, finding \Hnot$=68.8^{+5.0}_{-4.3}$\,\hubbunit\ and \Hnot$=62.3 \pm 9.1$\,\hubbunit\ respectively. These values are compatible at the $1\sigma$ level with our result of \result\,\hubbunit\ --- however, it is still useful to analyse differences in the methods, especially given that common data was used.

\Wu\ and \HS\ use 18 and 9 localised FRBs respectively, in both cases including a subset of the bursts used in this analysis, and those from other instruments. These include repeating FRBs which have been localised purely because they are repeaters, which presents a biased distribution of the underlying population \citep{Gardenier2019FRBPOPPY}, although the effect of this bias is difficult to determine. Furthermore, neither study accounts for observational biases against high DMs, which will systematically increase \Hnot\ by assuming the artificially low measured mean DMs reflect the true underlying distribution.

Both \Wu\ and \HS\ model DM contributions according to \eqref{eq:dmfrb}--\eqref{eq:dmlocal}, with \Wu\ in particular using very similar functional forms. \HS\ however uses Gaussian distributions in linear space to model both \dmhost\ and \dmcosmic; given that they are symmetric about the mean, they do not include the high-DM tail, nor the relatively sharp lower limit to DM for a given redshift, expected for \dmeg.

Importantly, both \Wu\ and \HS\ allow for an uncertainty in \dmlocal, using $\sigma_{\rm MW}=30$\,\dmunits, with \Wu\ also allowing \dmhalo\ to vary in the range 50--80\,\dmunits. Allowing for such uncertainties is an important next step in our model, due to the influence of \effect.

Neither \Wu\ nor \HS\ however allow assumed values for other parameters in their model to vary --- and \Hnot\ is particularly sensitive to the assumed value of \muhost\ (see \figref{fig:2DCorrelations}). \HS\ assumes \muhost=100\dmunits, while \Wu\ use a model for \muhost\ based on the simulations of \citet{Zhang2020_IllustrisTNG}, with \muhost\ increasing with $z$, and varying according to the class of FRB host galaxy (see discussion below). However, all values of \muhost\ are lower than our best-fit median value of $10^{\mu_{\rm host}}=186$\,\dmunits. Since their assumed values are low, their assumed \dmigm\ will be increased, which we attribute as being primarily responsible for the lower \Hnot\ values estimated by these authors.

\subsection{Sources of Uncertainty and Bias}
\label{sec:uncertainties}

In this analysis, we have not allowed for uncertainties in the cosmological parameters \fdz, \omegabhh, and the feedback parameter $F$. Of these, the current experimental uncertainty in \omegabhh\ is $\mathcal{O}\sim0.5$\% \citep{Mossa2020_Deuterium}, and negligible compared to errors in the current calculation. In the future, we will need to  marginalise over this uncertainty.

\begin{figure}
    \centering
    \includegraphics[width=3in]{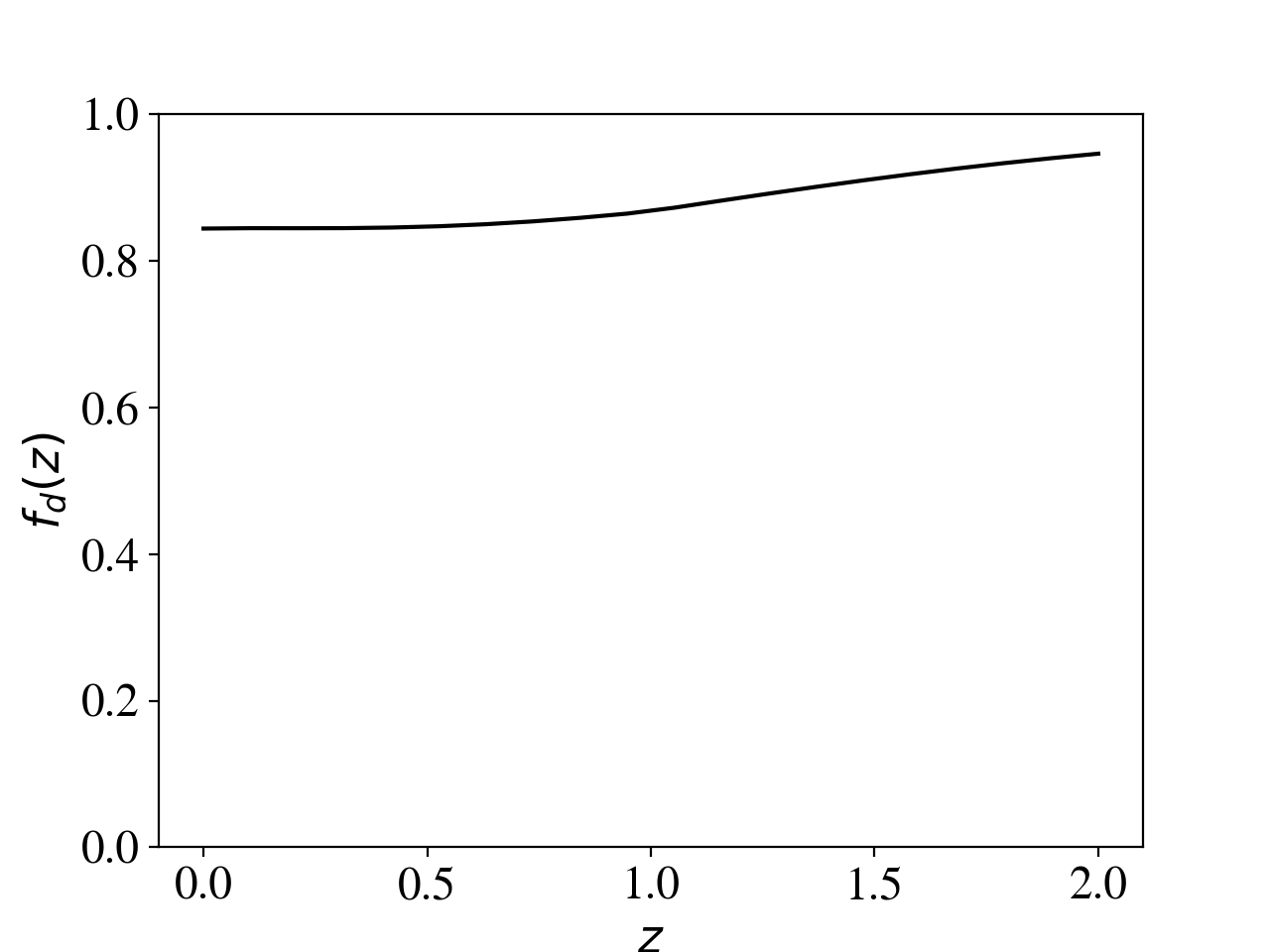}
    \caption{Modelled redshift evolution of the fraction of baryons that are diffused and ionized, \fdz, calculated according to \citet{Macquart2020} using code from \citet{frb}.}
    \label{fig:fdz}
\end{figure}

Our adopted estimate for the fraction of baryons that are diffuse and ionized, \fdz,
follows the methodology introduced in \citet{frb}, and discussed 
in further detail in \citet{Macquart2020}.
The approach uses the estimated mass density of baryons in dense (i.e.\ neutral)
gas and compact objects (stars, stellar remnants) from observations and
stellar population modelling.
For the present-day, one recovers $\fdzo=0.844$ using the \cite{PlanckCosmology2018}
parameters to estimate the total mass density in baryons. The redshift evolution of \fdz\ is plotted in \figref{fig:fdz}.
Regarding uncertainty in \fdz, the stellar mass estimate dominates
primarily through our imprecise knowledge of the stellar initial mass function (IMF).
If we assume a $30\%$ uncertainty in the stellar mass density (and associated
remnants), this translates to a $4\%$ uncertainty in \fdzo.
As the statistical power of the FRB sample grows, this systematic 
error in \fdz\ will rise in importance for \Hnot\ analysis.
It is possible, however, that upcoming experiments including weak-lensing
surveys will reduce the IMF uncertainties.
We also emphasize that at higher redshifts, $f_d$ increases as the masses
of galaxies decreases.  For example, a $30\%$ error in the stellar mass 
contribution at $z=1$ leads to a $\approx 2.5\%$ error in \fdzn.
In this respect, by including FRBs at $z>1$ one
can partially alleviate the uncertainty in \fdz.

The feedback parameter $F$ acts to smear the distribution of \dmigm\ about the mean, with smaller values representing a larger `feedback' effect 
that reduces the gas content of galactic halos, 
and hence a less clumpy Universe with a less smeared distribution of \dmigm\ \citep{Cen2006feedback}. To first order, the influence of $F$ is similar to the smearing of \dmhost\ by \sigmahost. Since both contribute to variance in \dmeg, and we fit \sigmahost\ to this, first-order uncertainties in $F$ are absorbed into \sigmahost. 
However, variance in \dmeg\ due to \dmhost\ is modelled as decreasing with $z$, while the 
absolute variance in \dmigm\ (due to $F$) increases with $z$. 
However, for future FRB samples --- particularly those including localisations at $z\gtrsim 1$ --- these terms should be separated, and $F$ explicitly fitted. For now, we note that \sigmahost\ and \Hnot\ do show significant (anti-)correlation, and thus potentially errors in our (somewhat arbitrarily) adopted value of $F=0.32$ could influence our measurement of \Hnot, although we expect that the main effect of such an error is to shift the fitted value of \sigmahost.

Related to this, we have used a fixed functional form for \dmhost, which while allowing for the reduced DM due to redshift, does not allow for evolution of the host galaxy properties themselves. Indeed, our adopted log-normal distribution is not theoretically motivated, but rather is a qualitatively good description of the expected high-DM tail of \dmhost. An improved model, including redshift evolution, will likely require combining FRB host galaxy studies \citep[e.g.][]{Bhandari+22} to derive
empirical correlations similar to treatments
of Hubble residuals with supernovae \citep[e.g.][]{Phillips93}. 
For instance, \citet{Zhang2020_IllustrisTNG} find a cosmic evolution of $\muhost = C_{\rm DM} (1+z)^{\alpha_{\rm DM}}$, with constant $33 \le C_{\rm DM} \le 96$\,\dmunits\ and $0.83 \le \alpha_{\rm DM} \le 1.08$, depending on galaxy type. Thus, when including the $(1+z)^{-1}$ redshift penalty to \dmhost, the observed distribution of \dmhost\ should remain almost constant with redshift. Since $C_{\rm DM}$ varies with galaxy type however, so that the distribution of galaxy types --- and thus presumably FRB host galaxies and thus \muhost\ --- will also vary with redshift. We suggest this topic for future investigations.

`\dmhost' represents all contributions to DM that come from the FRB occurring in a non-random part of the Universe, and thus includes the local cosmic structure (e.g.\ filament), halo, and interstellar medium of the host, and the immediate environment of the progenitor. Improvements in modelling should target all these aspects. One path forward is to use optical observations of the FRB host galaxy environment and intervening matter \citep{FLIMFLAM2022}, combined with properties such as the rotation measure and scattering of the FRB itself \citep{Cordes2022}, to constrain these values on a per-FRB basis.

Finally, we note that our models for the FRB luminosity function, and spectral behaviour, are relatively simplistic --- the true behaviour is likely more complicated. These are discussed in more detail in \james, and both can be investigated through near-Universe observations of FRBs.

\subsection{Comparison with other methods}

It is interesting to compare and contrast FRBs with other probes of \Hnot\ in the local Universe. The traditional method of using Type Ia Supernovae (SNIa) relies on these being calibratable standard candles, using the cosmological distance ladder and in particular Cepheid variables \citep[e.g.][ and references therein]{Riess2021}. This is certainly not the case for FRBs, which show a vast range of luminosity \citep{Spitler2014,Shannonetal2018}. However, using early Universe constraints on \omegabhh\ allows FRB DM to effectively become a `standard candle', directly relating the value of DM to distance via the Macquart relation. This means however that FRB measures of \Hnot\ in the local Universe will not be fully independent of early Universe cosmological fits,
but can nonetheless be used to identify an inconsistent cosmology.

Both methods suffer from potential biases due to a changing nature of host galaxies with redshift, and concerns regarding dust extinction biases in SNIa \citep[see e.g.][]{Sullivan2003_H0} are analogous to FRB detection biases against high DMs induced by FRB host galaxies. As discussed above, FRBs are sensitive to local Universe contributions to DM, which are less well-known than Galactic extinction effects on SN1a. Thus in many ways, FRB determination of \Hnot\ is qualitatively similar to, but independent of, measures derived from SNIa --- making them a perfect method for determining if the current Hubble tension is a result of systematic errors in the distance ladder, or a sign of new cosmology beyond $\Lambda$CDM.

Current limits on \Hnot\ from FRBs are much less accurate than those from SNIa, with statistical errors of ${}_{-8}^{+12}$\,\hubbunit\ compared to $\pm 1.01$\,\hubbunit\ for SN1a \citep{Riess2021}. New experiments promise to dramatically increase the rate of both samples --- the Legacy Survey of Space and Time (LSST) to be performed by the Rubin Observatory is expected to detect 400,000 Type Ia supernovae \citep{LSST2019}, while CRACO, DSA2000 \citep{DSA2000}, and CHORD \citep{CHORD} expect to increase the number of localised FRBs to the tens of thousands, although making use of this sample will require 
significant investment 
on optical telescopes to obtain 
host redshifts.

Gravitational wave (GW) detections also promise to constrain \Hnot\ \citep{Schutz1986}. Since the intrinsic signal strength is precisely predicted from general relativity, there is essentially no calibration uncertainty. The challenge however lies in the event rate of localisable GW signals, which is very low, with so-far only a single event yielding an uncertainty of $70.3_{-5.0}^{+5.3}$\,\hubbunit\ \citep{Hotokezaka2019_H0}. There are also only moderate improvements expected in the near future \citep{LIGO_2020_prospects}: a 1.8\% precision on \Hnot\ would require 50-100 binary neutron star (BNS) mergers with host galaxies, or 15 which also have afterglow information constraining their inclination angle \citep{Hotokezaka2019_H0}. The expected rate of BNS detection during the upcoming fourth observing run (`O4') of the LIGO--Virgo--KAGRA GW detectors,\footnote{https://www.ligo.caltech.edu/news/ligo20211115} at $7.7^{+11.9}_{-5.7}$\,yr$^{-1}$, is too uncertain to make hard predictions about the time required to reach these numbers \citep{Colombo2022}. The $\sim 78\%$ fraction of these events that will produce a kilonova (and hence redshift via host galaxy identification), and $\sim10\%$ fraction that will produce a detectable jet constraining their inclination angle, must also be considered \citep{Colombo2022}.

Therefore, provided that the systematic errors discussed in Section~\ref{sec:uncertainties} can be reduced, FRBs will be a valuable cosmological probe, should be at least as precise as GW-based approaches, and have the potential to approach SN1a in precision.

\section{Conclusions}

Using a sample of 16 localised and 60 unlocalised FRBs, we have fitted the observed values of \snr, $z$, and DM for each FRB, and the number of FRBs observed by each survey, using a Bayesian approach. We use the methodology of \citet{James2022Meth}, which includes the biasing effects of telescope beamshape and the FRB width, and models the FRB luminosity function, source evolution, and properties of the FRB population and their host galaxies. We have updated the method to allow \Hnot\ to vary while adjusting \omegab\ according to precise constraints on \omegabhh\ from the CMB. We find a best-fit value of \result\,\hubbunit, consistent with both direct and indirect measures of \Hnot\ \citep{SNOWMASScosmology}, and other estimates using FRBs \citep{WuH02022,HagstotzH02022}. This estimate was obtained with uniform priors over the entire range of \muhost, \sigmahost, \lemax, $\gamma$, and \sfrn\ where the likelihood is non-vanishing; and the plausible range of $-2 \le \alpha \le 0$ allowed by other studies.

We discuss systematic differences in the different methodologies for inferring \Hnot\ from FRB samples, and we attribute the lower values of \Hnot\ found by those studies to be primarily due to their low assumed values of \dmhost, which we fit, 
finding a median host contribution of $186^{+59}_{-48}$\,\dmunits\
(from an assumed Galactic halo contribution of $\dmhalo = 50\,\dmunits$).

The addition of new data confirms the previous result of \citet{James2022Lett} that the FRB population evolves with redshift in a manner consistent with the star-formation rate, and excludes no source evolution at $3\,\sigma$. This is consistent with young magnetar scenarios, and the presence of FRBs near spiral arms \cite[][]{Manningsspirals},  although we do not specifically exclude older progenitors. We have also constrained the frequency dependence of the FRB rate, albeit weakly, finding that $R_{\rm FRB} \propto \nu^{-1 \pm 0.85}$. Our estimated slope of the cumulative luminosity function is $\gamma=-0.95_{-0.15}^{+0.18}$, slightly flatter than previous estimates, and consistent with values derived from individual repeating FRBs.

We have used Monte Carlo simulations to predict that 100 localised FRBs from the first year of operation of the CRACO system of ASKAP will be able to constrain \Hnot\ with a statistical uncertainty 
of $\approx \pm 2.45 \hubbunit$, giving significant power to discriminate between different existing estimates. This motivates further work to address systematic uncertainties in our modelling, in particular to constrain or fit the fraction of cosmic baryons in diffuse ionized gas, and the contribution of the Milky Way halo to DM.

\section*{Acknowledgements}

Authors J.X.P., A.C.G., and N.T.\ as members 
of the Fast and Fortunate for FRB
Follow-up team, acknowledge support from 
NSF grants AST-1911140 and AST-1910471. This research was partially supported by the Australian Government through the Australian Research Council's Discovery Projects funding scheme (project DP210102103).
R.M.S.\ acknowledges support through Australian Research Council Future Fellowship FT190100155. S.B.\ is supported by a Dutch Research Council (NWO) Veni Fellowship (VI.Veni.212.058). The authors thank Evan Keane for comments on the manuscript.

This research has made use of the NASA/IPAC Extragalactic Database (NED),
which is operated by the Jet Propulsion Laboratory, California Institute of Technology,
under contract with the National Aeronautics and Space Administration. This research made use of Python libraries \textsc{Matplotlib} \citep{Matplotlib2007}, \textsc{NumPy} \citep{Numpy2011}, and \textsc{SciPy} \citep{SciPy2019}. This work was performed on the gSTAR national facility at Swinburne University of Technology. gSTAR is funded by Swinburne and the Australian Government’s Education Investment Fund. This work was supported by resources provided by the Pawsey Supercomputing Centre with funding from the Australian Government and the Government of Western Australia.

The Australian SKA Pathfinder is part of the Australia Telescope National Facility (https://ror.org/05qajvd42) which is managed by CSIRO. Operation of ASKAP is funded by the Australian Government with support from the National Collaborative Research Infrastructure Strategy. ASKAP uses the resources of the Pawsey Supercomputing Centre. Establishment of ASKAP, the Murchison Radio-astronomy Observatory and the Pawsey Supercomputing Centre are initiatives of the Australian Government, with support from the Government of Western Australia and the Science and Industry Endowment Fund. We acknowledge the Wajarri Yamatji people as the traditional owners of the Observatory site.

This research is based on observations collected at the European Southern Observatory under ESO programmes 0102.A-0450(A), 0103.A-0101(A), 0103.A-0101(B), 105.204W.001, 105.204W.002, and 105.204W.003.

\section*{Data Availability}

The code used in this study is available from GitHub \citep{frb,zdm}, which also includes the input FRB data. The calculated values of \pn\ and \pszdm\ for each dataset takes up several GB, and a download link is available upon request to the authors.


\bibliographystyle{mnras}
\bibliography{bibliography}

\appendix

\input{systematics_appendix}


\bsp	
\label{lastpage}
\end{document}

%% file: Tables/tab_model_params.tex
\begin{table*}
\centering
\begin{minipage}{170mm} 
\caption{Fiducial Set of Model Parameters. Parameters labelled with a * are re-fit as part of this work. 
\label{tab:param}}
\begin{tabular}{cccl}
\hline 
Parameter & Fiducial Value & Unit & Description 
\\ 
\hline 
$\log_{10} \mu_{s}$& 0.7& ms& Mean of log$_{10}$-scattering distribution at 600\,MHz\\ 
$\log_{10} \sigma_{s}$& 1.9& ms&  Standard deviation of log$_{10}$-scattering distribution at 600\,MHz \\ 
$\mu_{w}$& 1.70267& ms& $\log_{10}$ mean of intrinsic width distribution in ms\\ 
$\sigma_{w}$& 0.899148& ms& $\log_{10}$ sigma of intrinsic width distribution in ms\\ 
DM$_{\rm ISM}$ & NE2001 & \dmunits & DM for the Milky Way Interstellar Medium \\ 
${\rm DM}_{\rm halo}$& 50& pc cm$^{-3}$& DM for the Galactic halo\\ 
$n_{\rm sfr}^*$& 0.73& & Scaling of FRB rate density with star-formation rate\\ 
$H_0^*$& 67.66& km s$^{-1}$ Mpc$^{-1}$& Hubble's constant\\ 
$\Omega_{\Lambda}$& 0.68885& & Dark energy / cosmological constant (in current epoch)\\ 
$\Omega_m$& 0.30966& & Matter density in current epoch\\ 
$\Omega_b$& 0.04897& & Baryon density in current epoch\\ 
$\Omega_b h^2$& 0.02242& & Baryon density weighted by $h_{100}^2$\\ 
$\mu_{\rm host}^*$& 2.18& & $\log_{10}$ mean of DM host contribution in pc cm$^{-3}$\\ 
$\sigma_{\rm host}^*$& 0.48& & $\log_{10}$ sigma of DM host contribution in pc cm$^{-3}$\\ 
$f_d(z=0)$ & 0.844 & & Fraction of baryons that are diffuse and ionized at $z=0$ \\ 
$F$& 0.32& & F parameter in DM$_{\rm cosmic}$ PDF for the Cosmic web\\ 
$\log_{10} E_{\rm min}$& 30& erg& $\log_{10}$ of minimum FRB energy \\ 
$\log_{10} E_{\rm max}^*$& 41.4& erg& $\log_{10}$ of maximum FRB energy\\ 
$\alpha^*$& 0.65& & Power-law index of frequency dependent FRB rate, $R \sim \nu^\alpha$\\ 
$\gamma^*$& -1.01& & Slope of luminosity distribution function\\ 
\hline 
\end{tabular} 
\end{minipage} 
\end{table*}

%% file: Tables/tab_ics_frbs.tex
\begin{table*}
\centering
\begin{minipage}{170mm} 
\centering
\caption{ASKAP incoherent sum FRBs used in this analysis.
Given is the FRB name, SNR-maximising DM, \dmism\ estimated using the NE2001 model of \citet{CordesLazio01}, central frequency of observation $\nu$, measured signal-to-noise ratio SNR, redshift $z$, posterior probability of host associations $\pox$, and original reference. Where redshifts are not given, this is because (a): no voltage data were dumped, preventing radio localization; (b) optical follow-up observations are not yet complete; 
(c) Substantial Galactic extinction has 
challenged follow-up optical observations; 
(d) the host galaxy appears too distant to accurately measure a redshift. \label{tab:ics_frbs}
}
\begin{tabular}{cccccccl}
\hline 
Name & DM & \dmism & $\nu$ & SNR & $z$ & $\pox$ & Ref.\ \\
& (\dmunits) & (\dmunits) & (MHz) & &  & &
\\ 
\hline 
\multicolumn{7}{c}{\icslow} \\
\hline
20191001 & 506.92 & 44.2  & 919.5 & 62.0 & 0.23    & 0.973 & \citet{Bhandari2020b} \\
20200430 & 380.1  & 27.0    & 864.5 & 16.0 & 0.161   & 1.000 & \citet{Heintz2020} \\
20200906 & 577.8  & 35.9  & 864.5 & 19.2 & 0.36879 & 1.000 & \citet{Bhandari+22} \\
20210807 & 251.9  & 121.2 & 920.5 & 47.1 & 0.12927 & 0.957 & \spirals  \\
\hline
20200627 & 294.0    & 40.0    & 920.5 & 11.0 & (a)      & n/a & \multirow{4}{*}{\icspaper} \\
20210320 & 384.8  & 42.2  & 864.5 & 15.3 & 0.2797    & 0.999 &  \\
20210809 & 651.5  & 190.1 & 920.5 & 16.8 & (a)      & n/a & \\
20211203 & 636.2  & 63.4  & 920.5 & 14.2 & (b)      & n/a &  \\
\hline 
\multicolumn{7}{c}{\icsmid} \\
\hline
20180924   & 362.4  & 40.5 & 1297.5 & 21.1 & 0.3214 & 0.999 &\citet{Bannister2019} \\
20181112   & 589.0  & 40.2 & 1297.5 & 19.3 & 0.4755 & 0.927 &\citet{Prochaska2019} \\
\hline
20190102   & 364.5  & 57.3 & 1271.5 & 14 & 0.291 & 1.000 & \multirow{4}{*}{\citet{Macquart2020}} \\
20190608   & 339.5  & 37.2 & 1271.5 & 16.1 & 0.1178 & 1.000 & \\
20190611.2 & 322.2  & 57.6 & 1271.5 & 9.3  & 0.378 & 0.980 & \\
20190711   & 594.6  & 56.6 & 1271.5 & 23.8 & 0.522 & 0.999 & \\ 
\hline
20190714   & 504.7  & 38.5 & 1271.5 & 10.7 & 0.209 & 1.000 & \citet{Heintz2020} \\
20191228   & 297.5  & 32.9 & 1271.5 & 22.9 & 0.243 & 1.000 & \citet{Bhandari+22} \\
20210117   & 730    & 34.4 & 1271.5 & 27.1 & 0.2145 & 0.999 & \bhandariinprep  \\
\hline
20210214   & 398.3  & 31.9 & 1271.5 & 11.6 & (a) & n/a & \multirow{3}{*}{\icspaper} \\
20210407   & 1785.3 & 154  & 1271.5 & 19.1 & (c) & n/a &  \\
20210912   & 1234.5 & 30.9 & 1271.5 & 31.7 & (d) & n/a & \\
\hline
20211127   & 234.83 & 42.5 & 1271.5 & 37.9 &  0.0469 & 0.998 & \spirals \\
\hline 
\multicolumn{7}{c}{\icshigh} \\
\hline
20211212 & 206 & 27.1 & 1632.5 & 12.8 & 0.0715 & 0.998 & \spirals \\
\end{tabular} 
\end{minipage} 
\end{table*}

%% file: Tables/tab_fe_pks_frbs.tex

\begin{table*}
\centering
\caption{Properties of \fe\ and \mb\ FRBs previously excluded due to their relatively high \dmism, which are now included in this analysis. Given is the original FRB designation; measured total DM and DM$_{\rm ISM}$ estimated by the NE2001 model \citep{CordesLazio01} in pc\,cm$^{-3}$, and ratio of measured to threshold SNR.} \label{tab:parkes_frbs}
\begin{tabular}{c c c c c}
\hline
FRB & DM & DM$_{\rm ISM}$ & SNR & Ref.\ \\
\hline
\multicolumn{5}{c}{\mb} \\
\hline
20150610 & 1593.9 & 104 & 18 & \multirow{2}{*}{\citet{Bhandarietal2018}}\\
20151206 & 1909.8  & 239 & 10 & \\
\hline
20171209 & 1457.4  & 329 & 40 &  \multirow{2}{*}{\citet{Oslowski2019}}\\
20180714 & 1467.92 & 254 & 22 & \\
\hline
20150418 & 776.2 & 164 & 39 & \citet{keane2016host}\\
20010125 & 790     & 105 & 17 & \citet{Burke2014_011025} \\
20010621 & 745     & 502 & 16.3 & \citet{Keane2011} \\
20150215 & 1105.6  & 405 & 19 & \citet{Petroff2017_150215}\\
\hline
\multicolumn{5}{c}{\fe} \\
\hline
20180315 & 479.0 & 100.8 & 10.5 & \citet{Macquart2019a} \\
20180430 & 264.1 & 169 & 28.2 & \citet{Qiu2019} \\
\hline
\end{tabular}
\end{table*}

%% file: Tables/tab_cube_params.tex
\begin{table}
    \centering
    \begin{tabular}{c|c c c}
      Parameter   &  Min & Max & Increment \\
      \hline
     \Hnot      & 55  & 101  & 2  \\
     \lemax     & 40.5  & 42.5  & 0.1   \\
     $\alpha$    & 0  & 2  & 0.5  \\
     $\gamma$   & -1.5  & -0.5  & 0.1  \\
     $n$        & 0  & 3  & 0.25  \\
     \muhost    & 1.5  & 2.6  & 0.1   \\
     \sigmahost & 0.3  & 1.1  & 0.1  \\
    \end{tabular}
    \caption{Parameters and their values (in linear increments from min to max values) at which the joint likelihood \pszdm\ was evaluated. \lemin\ was fixed at $10^{30}$\,erg.}
    \label{tab:cube_params}
\end{table}

%% file: Tables/tab_extended_results.tex
\begin{table}
\renewcommand{\arraystretch}{1.5}
    \centering
    \caption{Best-fitting parameter values and associated confidence intervals for each fitted parameter. For parameters other than \Hnot, limits are given for different priors on \Hnot: the `standard' prior, covering both early- and local-Universe measurements of \Hnot\ (see text); fixing \Hnot\ to 67.4\,\hubbunit\ and 73.04\,\hubbunit; and a flat prior between 55 and 101\,\hubbunit. For $\alpha$, only approximate 68 per cent errors are given (see text).}
    \label{tab:confidence_intervals}
     \begin{tabular}{c|c c c c c c}
     & & & 68 per & 90 per & 95 per & 99.7 per \\
     Parameter & Prior & Best Fit & cent  & cent  & cent & cent \\
     \hline
      $H_0$  & N/A & 73.0 & $_{-8}^{+12}$ & $_{-12}^{+22}$ & $_{-13}^{+29}$ & $_{-17}^{+48}$  \\
      \hline
$\log_{10} E_{\rm max}$ & Std & $41.26$& $_{-0.22}^{+0.27}$ & $_{-0.33}^{+0.50}$ & $_{-0.38}^{+0.64}$ & $_{-0.51}^{+1.02}$ \\
    & Flat & $41.20$& $_{-0.25}^{+0.29}$ & $_{-0.39}^{+0.52}$ & $_{-0.45}^{+0.67}$ & $_{-0.61}^{+1.06}$ \\
    & 73.04 & $41.21$& $_{-0.21}^{+0.26}$ & $_{-0.32}^{+0.49}$ & $_{-0.37}^{+0.63}$ & $_{-0.49}^{+1.03}$ \\
    & 67.4 & $41.33$& $_{-0.22}^{+0.27}$ & $_{-0.33}^{+0.50}$ & $_{-0.38}^{+0.63}$ & $_{-0.50}^{+0.99}$ \\
$\alpha$ & Std & $-0.99$& $_{-1.01}^{+0.99}$ & \multicolumn{3}{c}{N/A} \\
    & Flat & $-0.92$& $_{-1.08}^{+0.92}$ & \multicolumn{3}{c}{N/A} \\
    & 73.04 & $-0.95$& $_{-1.05}^{+0.95}$ & \multicolumn{3}{c}{N/A} \\
    & 67.4 & $-1.03$& $_{-0.97}^{+1.03}$ & \multicolumn{3}{c}{N/A} \\
$\gamma$ & Std & $-0.95$& $_{-0.15}^{+0.18}$ & $_{-0.23}^{+0.30}$ & $_{-0.27}^{+0.36}$ & $_{-0.36}^{+0.45}$ \\
    & Flat & $-0.94$& $_{-0.15}^{+0.18}$ & $_{-0.24}^{+0.30}$ & $_{-0.27}^{+0.37}$ & $_{-0.36}^{+0.44}$ \\
    & 73.04 & $-0.95$& $_{-0.15}^{+0.18}$ & $_{-0.23}^{+0.30}$ & $_{-0.27}^{+0.37}$ & $_{-0.36}^{+0.45}$ \\
    & 67.4 & $-0.95$& $_{-0.15}^{+0.17}$ & $_{-0.23}^{+0.29}$ & $_{-0.27}^{+0.36}$ & $_{-0.35}^{+0.45}$ \\
$n_{\rm sfr}$ & Std & $1.13$& $_{-0.41}^{+0.49}$ & $_{-0.65}^{+0.77}$ & $_{-0.77}^{+0.90}$ & $_{-1.09}^{+1.19}$ \\
    & Flat & $1.08$& $_{-0.41}^{+0.50}$ & $_{-0.64}^{+0.78}$ & $_{-0.76}^{+0.92}$ & $_{-1.05}^{+1.21}$ \\
    & 73.04 & $1.10$& $_{-0.41}^{+0.50}$ & $_{-0.63}^{+0.78}$ & $_{-0.75}^{+0.92}$ & $_{-1.06}^{+1.21}$ \\
    & 67.4 & $1.15$& $_{-0.41}^{+0.49}$ & $_{-0.66}^{+0.76}$ & $_{-0.79}^{+0.89}$ & $_{-1.11}^{+1.18}$ \\
$\mu_{\rm host}$ & Std & $2.27$& $_{-0.13}^{+0.12}$ & $_{-0.23}^{+0.21}$ & $_{-0.28}^{+0.26}$ & $_{-0.47}^{+0.33}$ \\
    & Flat & $2.33$& $_{-0.14}^{+0.13}$ & $_{-0.25}^{+0.21}$ & $_{-0.31}^{+0.25}$ & $_{-0.57}^{+0.27}$ \\
    & 73.04 & $2.30$& $_{-0.12}^{+0.11}$ & $_{-0.20}^{+0.19}$ & $_{-0.25}^{+0.24}$ & $_{-0.40}^{+0.30}$ \\
    & 67.4 & $2.23$& $_{-0.14}^{+0.13}$ & $_{-0.25}^{+0.22}$ & $_{-0.31}^{+0.28}$ & $_{-0.51}^{+0.37}$ \\
$\sigma_{\rm host}$ & Std & $0.55$& $_{-0.09}^{+0.12}$ & $_{-0.13}^{+0.22}$ & $_{-0.16}^{+0.30}$ & $_{-0.21}^{+0.51}$ \\
    & Flat & $0.53$& $_{-0.08}^{+0.11}$ & $_{-0.13}^{+0.22}$ & $_{-0.15}^{+0.29}$ & $_{-0.20}^{+0.52}$ \\
    & 73.04 & $0.54$& $_{-0.08}^{+0.11}$ & $_{-0.13}^{+0.20}$ & $_{-0.15}^{+0.27}$ & $_{-0.20}^{+0.48}$ \\
    & 67.4 & $0.57$& $_{-0.09}^{+0.13}$ & $_{-0.14}^{+0.25}$ & $_{-0.17}^{+0.33}$ & $_{-0.21}^{+0.51}$ \\
    \hline
    \end{tabular}
\end{table}

%% file: Tables/tab_MC.tex
\begin{table}
\centering
\caption{Monte Carlo FRBs generated for the (in development) CRACO system on ASKAP (see Section \ref{sec:craco_mc_results}). \label{tab:MC}}
\begin{minipage}{170mm} 

\begin{tabular}{ccc}
\hline 
DM & \snr & $z$ 
\\ 
 & (\dmunits) 
\\ 
\hline 
186.7& 15.9& 0.15\\ 
1179.5& 26.0& 1.321\\ 
438.5& 20.2& 0.062\\ 
315.3& 17.9& 0.089\\ 
833.1& 10.5& 0.949\\ 
595.3& 32.8& 0.25\\ 
313.4& 16.8& 0.37\\ 
568.5& 13.1& 0.629\\ 
143.5& 11.0& 0.026\\ 
743.4& 12.6& 0.812\\ 
941.9& 10.0& 0.755\\ 
460.5& 10.7& 0.355\\ 
1271.8& 12.1& 0.432\\ 
1308.6& 15.7& 1.273\\ 
567.9& 12.4& 0.608\\ 
410.5& 84.7& 0.057\\ 
391.1& 117.3& 0.251\\ 
1287.8& 9.7& 1.592\\ 
634.8& 15.9& 0.378\\ 
383.0& 15.6& 0.176\\ 
372.7& 16.2& 0.167\\ 
237.3& 13.3& 0.041\\ 
478.8& 27.0& 0.354\\ 
835.0& 51.2& 0.735\\ 
282.6& 10.6& 0.263\\ 
151.4& 223.4& 0.046\\ 
819.9& 13.1& 0.75\\ 
162.3& 17.3& 0.138\\ 
371.1& 25.0& 0.282\\ 
357.4& 16.3& 0.262\\ 
331.8& 40.2& 0.085\\ 
557.5& 88.5& 0.289\\ 
818.5& 9.8& 0.66\\ 
1257.6& 22.4& 0.699\\ 
1116.3& 11.6& 1.367\\ 
2259.2& 12.3& 1.338\\ 
307.9& 22.6& 0.243\\ 
1311.1& 24.4& 1.712\\ 
848.2& 11.6& 1.006\\ 
1060.6& 10.6& 1.108\\ 
785.5& 10.0& 0.164\\ 
484.9& 11.7& 0.52\\ 
481.2& 9.6& 0.424\\ 
484.7& 14.9& 0.61\\ 
260.6& 12.9& 0.054\\ 
393.8& 12.7& 0.291\\ 
273.7& 22.7& 0.182\\ 
534.4& 21.0& 0.556\\ 
703.6& 10.8& 0.195\\ 
335.9& 18.5& 0.329\\
\hline 
\end{tabular}
\begin{tabular}{ccc}
\hline 
DM & \snr & $z$ 
\\ 
 & (\dmunits) 
\\ 
\hline 
898.1& 32.2& 0.718\\ 
582.2& 13.8& 0.571\\ 
636.2& 23.1& 0.441\\ 
735.7& 27.7& 0.482\\ 
1405.0& 11.4& 1.318\\ 
1083.0& 19.3& 1.101\\ 
709.4& 12.4& 0.58\\ 
1794.1& 14.9& 1.849\\ 
736.2& 11.0& 0.546\\ 
808.6& 33.4& 0.472\\ 
352.0& 16.0& 0.126\\ 
447.5& 38.3& 0.543\\ 
1346.2& 12.3& 1.567\\ 
428.3& 12.3& 0.537\\ 
421.8& 28.0& 0.018\\ 
602.3& 16.3& 0.093\\ 
1110.1& 94.0& 0.439\\ 
303.5& 10.8& 0.137\\ 
799.7& 18.8& 0.465\\ 
309.6& 30.0& 0.159\\ 
3446.6& 25.6& 0.08\\ 
721.5& 25.8& 0.306\\ 
296.3& 21.4& 0.134\\ 
573.2& 19.6& 0.464\\ 
184.5& 11.4& 0.063\\ 
580.0& 15.6& 0.632\\ 
754.0& 9.6& 0.103\\ 
391.9& 11.2& 0.43\\ 
282.0& 21.7& 0.073\\ 
548.7& 25.8& 0.35\\ 
449.7& 12.6& 0.019\\ 
187.9& 40.8& 0.052\\ 
522.1& 11.8& 0.353\\ 
233.3& 85.5& 0.081\\ 
923.9& 9.9& 0.961\\ 
568.1& 15.9& 0.295\\ 
1327.1& 18.6& 0.437\\ 
901.0& 39.8& 0.744\\ 
776.9& 14.5& 0.786\\ 
359.8& 46.4& 0.372\\ 
733.7& 18.7& 0.633\\ 
685.6& 9.8& 0.765\\ 
568.9& 14.8& 0.644\\ 
398.7& 23.3& 0.239\\ 
664.2& 15.8& 0.495\\ 
326.6& 14.2& 0.251\\ 
726.0& 23.7& 0.089\\ 
342.1& 48.2& 0.029\\ 
376.5& 47.9& 0.279\\ 
271.9& 18.9& 0.237\\ 
\hline 
\end{tabular} 
\end{minipage} 
\end{table}

%% file: systematics_appendix.tex
\section{Studies of systematic effects}
\label{sec:systematics}

In this Appendix, we present several studies of potential systematic effects which, if not properly controlled, could bias our estimates of \Hnot.

\subsection{Missing low-\snr\ FRBs}
\label{sec:snr_problem}

The question of completeness in FRB surveys has been discussed by several authors \citep[e.g.][]{Macquart2018a,Bhandarietal2018,Jamesetal2019a}. From a simulation perspective, we assume completeness above some signal-to-noise threshold, 
\snrth, i.e.\ all FRBs with ${\rm SNR}_{\rm FRB} > \snrth$ are detected. However, practically, this may not be case. FRBs which pass the threshold are usually required to be either visually inspected, and/or analysed by an algorithm, to distinguish these events from RFI. Such an inspection --- whether done by human or machine --- will be more likely to fail for weak FRBs than for strong ones. This observational bias has been suggested as one reason why the first FRB to be discovered, the `Lorimer burst' \citep[FRB 20010724;][]{Lorimer2007}, was exceptionally bright --- because all the other bursts which had been viewed, but were not so bright, had not been identified \citep{Macquart2018a}. It has also been suggested to explain the dearth of Parkes FRBs with $\snr < 14$, although there is no evidence of such a bias from the CRAFT/FE observations \citep{James2019a_source_counts}.

\begin{figure}
    \centering
    \includegraphics[width=3.5in]{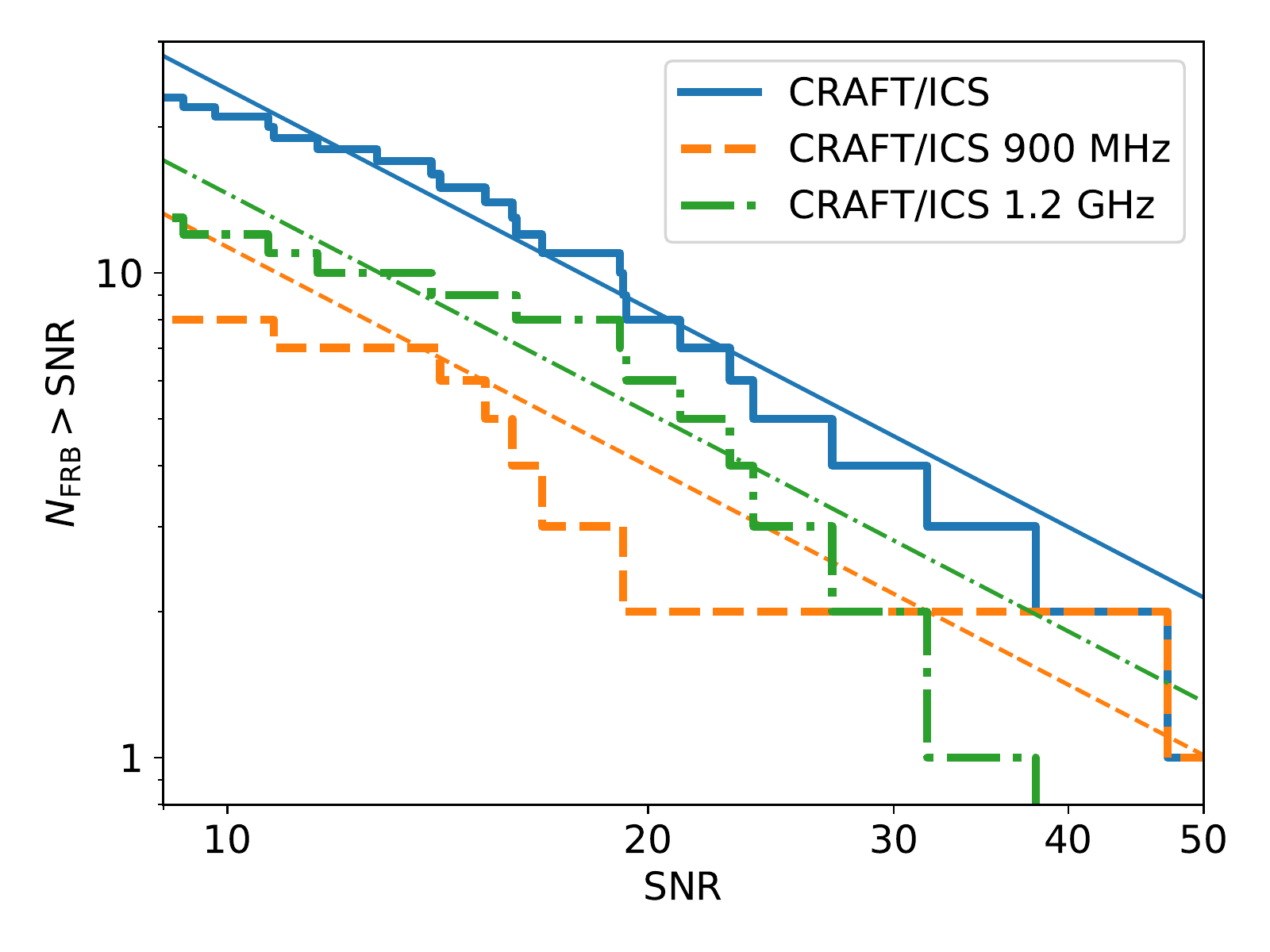}
    \caption{Cumulative event count as a function of FRB SNR for all \ics\ observations (blue, solid), and separated into \icslow\ (orange, dashed) and \icsmid\ (green, dot-dash). Only two \icshigh\ FRBs were detected and are not shown separately. Also shown are fits to a $\snr^{-1.5}$ power law.}
    \label{fig:logNlogS}
\end{figure}

For \ics\ observations, the detection threshold, \snrth, initially had to be set as high as 14$\sigma$ to reduce the number of false RFI candidates. Subsequently, an improved clustering algorithm, \textsc{snoopy2}, was developed to analyse all events satisfying ${\rm SNR}_{\rm FRB} > \snrth$ and reject RFI. This algorithm has been shown to pass all FRBs detected in ASKAP/FE observations. Furthermore, ASKAP/ICS observations of pulsars have not shown any evidence of behaviour that would reject true FRBs, i.e.\ all pulsar pulses above the detection threshold are not rejected. Thus \ics\ observations typically use a threshold of 9$\sigma$.

Figure~\ref{fig:logNlogS} plots the number of ASKAP FRBs observed above different \snr\ values. In a Euclidean Universe, this logN--logS curve should have a power-law slope of -1.5. However, for CRAFT/ICS, there is evidence for an inflection point near \snr\ of 15. While part of this inflection can be explained by early observations with a high \snrth, the threshold had stablised by the time of the 900\,MHz observations --- and the inflection point is present at both frequency ranges.

The source of this deficit of low-\snr\ FRBs is unknown. One potential solution is to artificially increase the detection threshold to an \snr\ of 15, and to discard all FRBs with $\snr < 15$. Before discarding valuable events however, we first investigate the potential bias of including FRBs in the range $9.5< \snr <15$.

\begin{figure}
    \centering
    \includegraphics[width=3.5in]{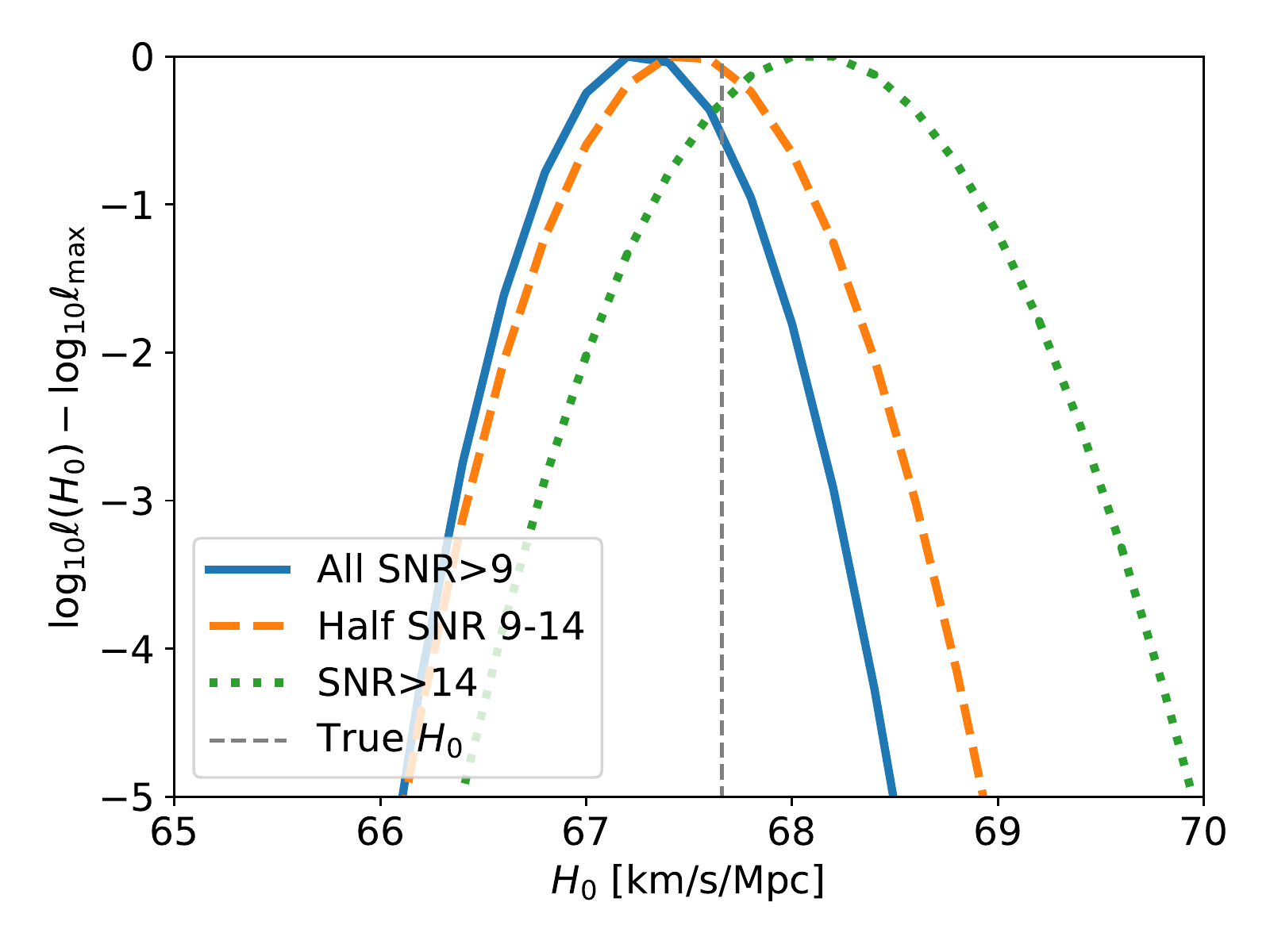}
    \caption{Simulated sensitivity to \Hnot\ of \icsmid\ observations when using all events with $\snr\ge9$, excluding half in the range $9\le\snr <  14$, and only using events with $\snr\ge14$.}
    \label{fig:snr_bias}
\end{figure}

To do so, we simulate 1000 FRBs from \icsmid\ using $\snrth =9$, and calculate 
the likelihood $\ell$(\Hnot) while holding 
all other parameters constant. As this is a single survey, \pn\ has no contribution, since the population density can always be appropriately scaled. The result is shown in Figure~\ref{fig:snr_bias}. We then repeat the calculation by randomly removing half of all FRBs in the range $9 \le \snr \le 14$ while keeping $\snrth =9$ as in the main analysis, and by removing all FRBs with $\snr < 15$ but accounting for this via setting $\snrth=15$. Thus both $\snr>9$ and $\snr>14$ calculations should present unbiased (but statistically fluctuating) measures of \Hnot, while the `half SNR 9--14' sample will present a biased estimate.

In Figure~\ref{fig:snr_bias}, the unbiased samples correctly reproduce the simulated true value of \Hnot\ with errors of $\pm 0.5$\,\hubbunit. The biased half 9--14 sample produces \Hnot\ in-between these two values --- very close to the simulated truth. Thus while we cannot exclude that missing FRBs in the range $9 \le {\rm SNR} \le 14$ results in a bias, this bias is smaller than the random deviation in \Hnot\ when using 1000 FRBs. Therefore we include the six FRBs with $9 \le {\rm SNR} \le 14$ in our sample, and use ${\rm SNR}_{\rm th}=9$; and we suggest that all near-future FRB surveys do the same.

\subsection{Effect of ISM}
\label{sec:ISM_effect}

The Milky Way interstellar medium (ISM) increases the dispersion measure of extragalactic FRBs at low Galactic latitudes through an increasing \dmism. This in turn reduces the sensitivity of FRB surveys at these latitudes in terms of extragalactic dispersion measure, \dmeg. Whether or not the apparent paucity of FRBs observed at low Galactic latitudes by Parkes \citep{Petroff2014a} is due to this effect or e.g.\ interstellar scintillation \citep{MacquartJohnston2015}, or is indeed even statistically significant \citep{Bhandarietal2018}, remains undetermined. Here, we model the effects of \dmism\ on the redshift distribution $z$ of observable FRBs using \icsmid. Note that the Galactic Plane also reduces the ability of optical follow-up observations to identify the FRB host galaxy, as discussed in \secref{sec:no_hosts}. Here we only consider only the effects on the initial detection of FRBs with radio waves.

\begin{figure}
    \centering
    \includegraphics[width=3.5in]{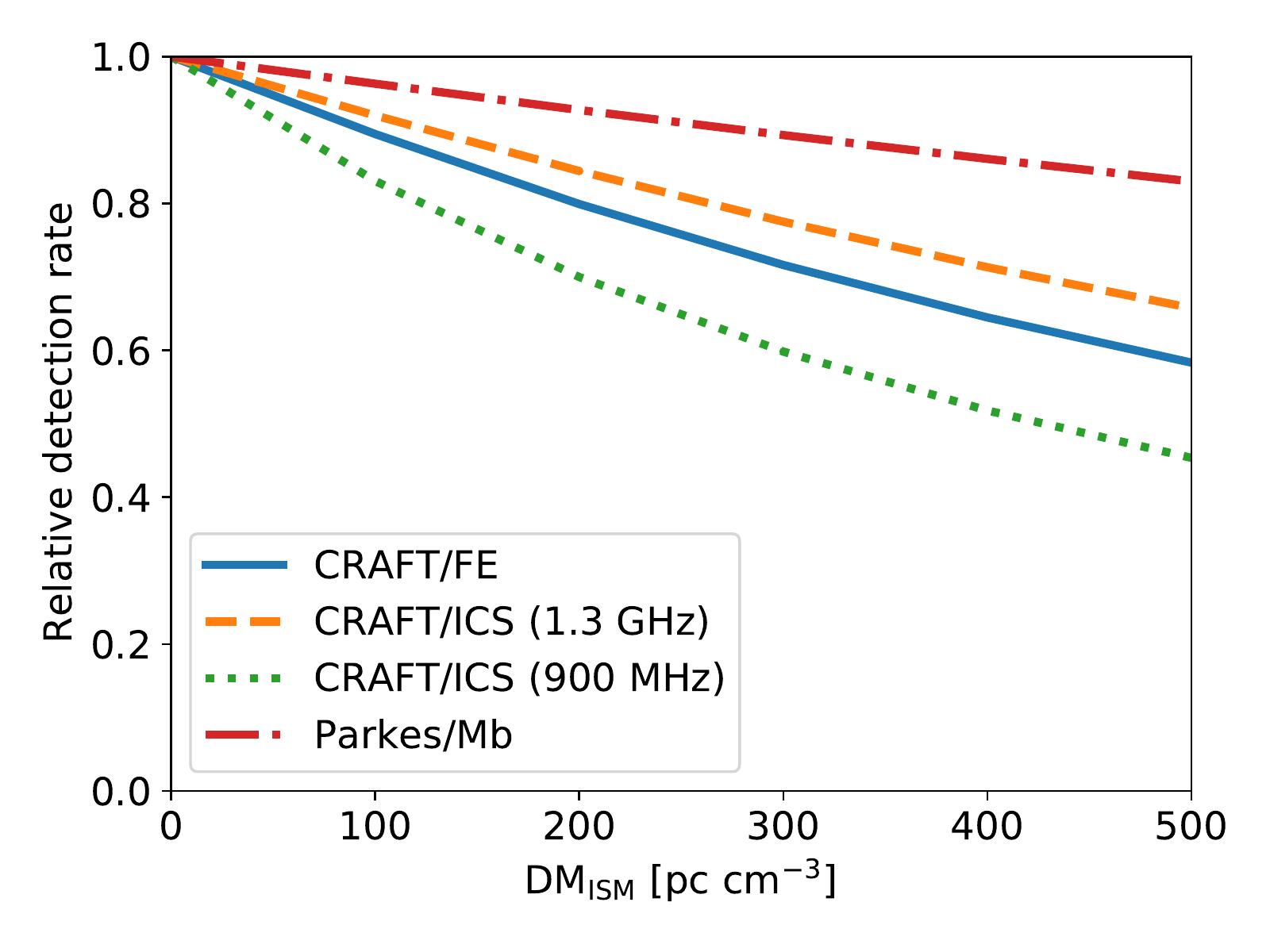}
    \caption{Dependence of total FRB detection rate on \dmism, relative to the hypothetical rate when \dmism$=0$.}
    \label{fig:dm_ism_rate}
\end{figure}

In \figref{fig:dm_ism_rate}, we plot the total (simulated) detectable FRB rate as a 
function of \dmism\ for the FRB surveys used in this work. In the range of 0--500 \pccc, the detection rate falls the least ($\sim$20\%) for \mb, due to its high frequency resolution negating the effects of increased DM smearing; and the most ($\sim$50\%) for \icslow\ due to its lower observation frequency and hence greater DM smearing.

\begin{figure}
    \centering
    \includegraphics[width=3.5in]{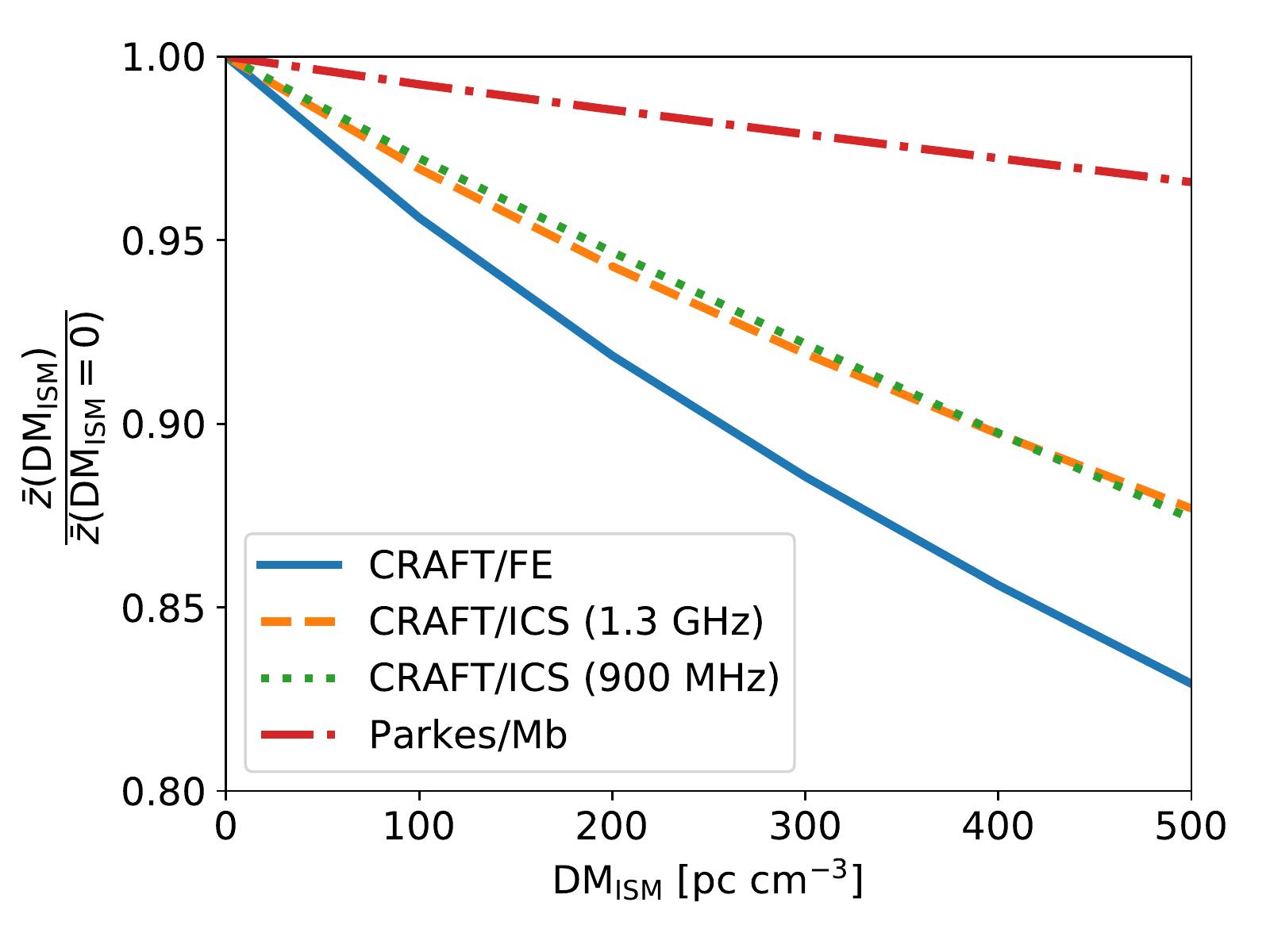}
    \caption{Dependence of mean detected redshift $z$ on ${\rm DM}_{\rm ISM}$.}
    \label{fig:dm_ism_mean_redshift}
\end{figure}

The effect of this reduced sensitivity on the observable redshift distribution is shown in \figref{fig:dm_ism_mean_redshift}. An increased \dmism\ decreases the mean redshift of detectable FRBs, $\bar{z}$, for all surveys considered here. Again, \mb\ is the least affected survey. However, the relative effect on \fe\ (17\% reduction in $\bar{z}$ at $\dmism=500\pccc$ relative to $\dmism=0\pccc$) is now greater than that on \icslow. This is likely because the lower sensitivity of \fe\ leads to, on-average, lower values of \dmeg, so that a moderate increase in \dmism\ has a proportionally greater effect on sensitivity and hence $\bar{z}$.

To test the effect of this approximation on \Hnot, we use simulated FRBs from \icsmid\ and the best-fit parameters of \citet{James2022Lett}.
We simulate four sets of FRBs, setting \dmism to $0$, $100$, $200$, and $500$ \pccc. To illustrate the bias effect of \dmism\ on \Hnot, we evaluate the likelihood $\ell$ by changing \Hnot\ only, subject to the constraint on \omegabhh.

\begin{figure}
    \centering
    \includegraphics[width=3.5in]{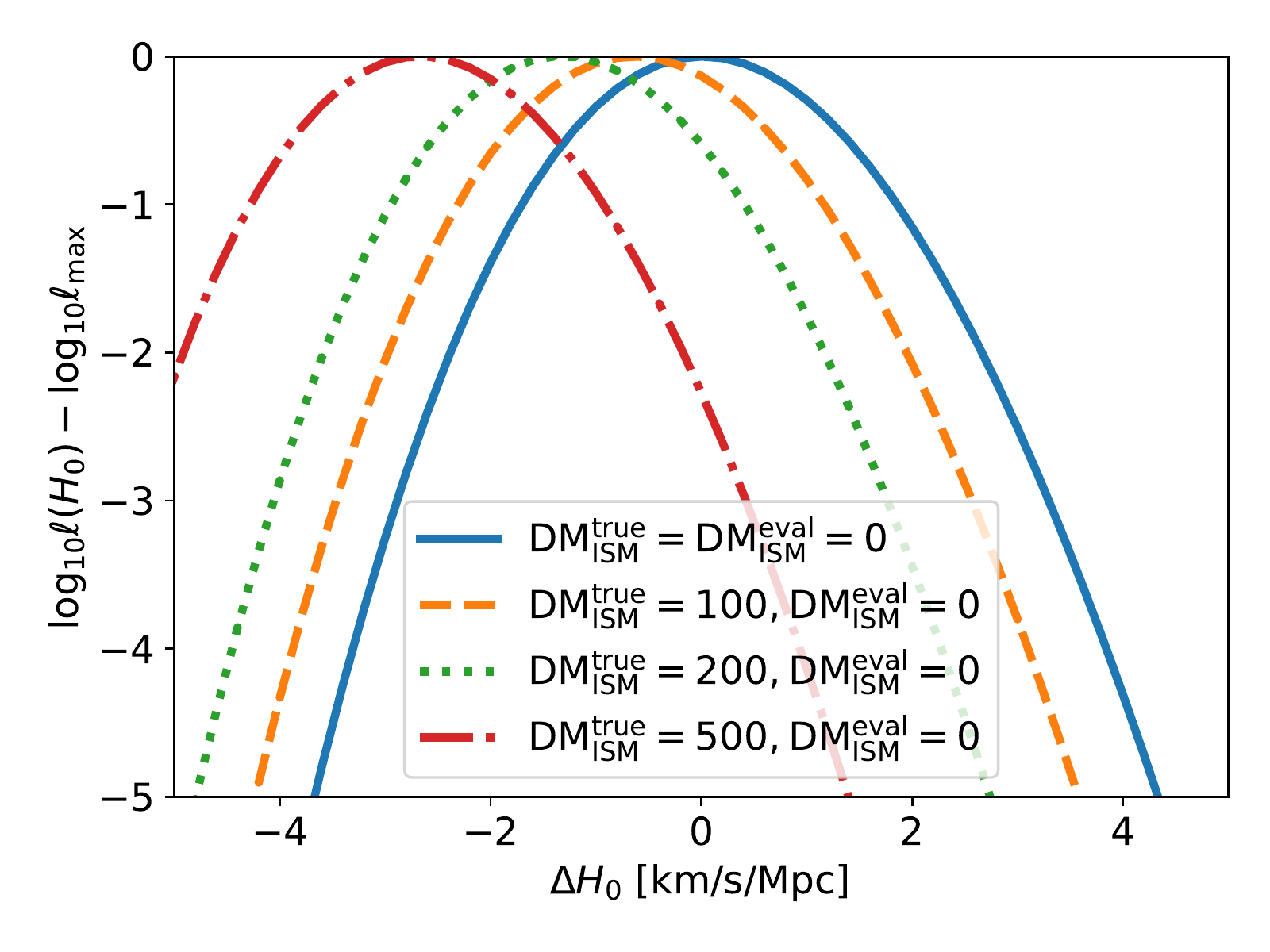}
    \caption{Effect on evaluated likelihoods of \Hnot\ (relative to maximum) when using an incorrect value of \dmism\ for evaluation.
    }
    \label{fig:dm_ism_effect}
\end{figure}

For our first test, we use samples of FRBs with \dmeg\ generated at true values of \dmism,  ${\rm DM}_{\rm ISM}^{\rm true}$, of $0$, $100$, $200$, and $500$ \pccc, but \Hnot\ is evaluated assuming a different value, ${\rm DM}_{\rm ISM}^{\rm eval}=0$. The results are shown in \figref{fig:dm_ism_effect}. In all cases, the most likely value of \Hnot\ is evaluated as being lower, by 0.6, 1.4, and 2.6 \hubbunit for $100$, $200$, and $500$ \pccc respectively, to account for seeing FRBs with lower \dmfrb\ for a given redshift.

\begin{figure}
    \centering
    \includegraphics[width=3.5in]{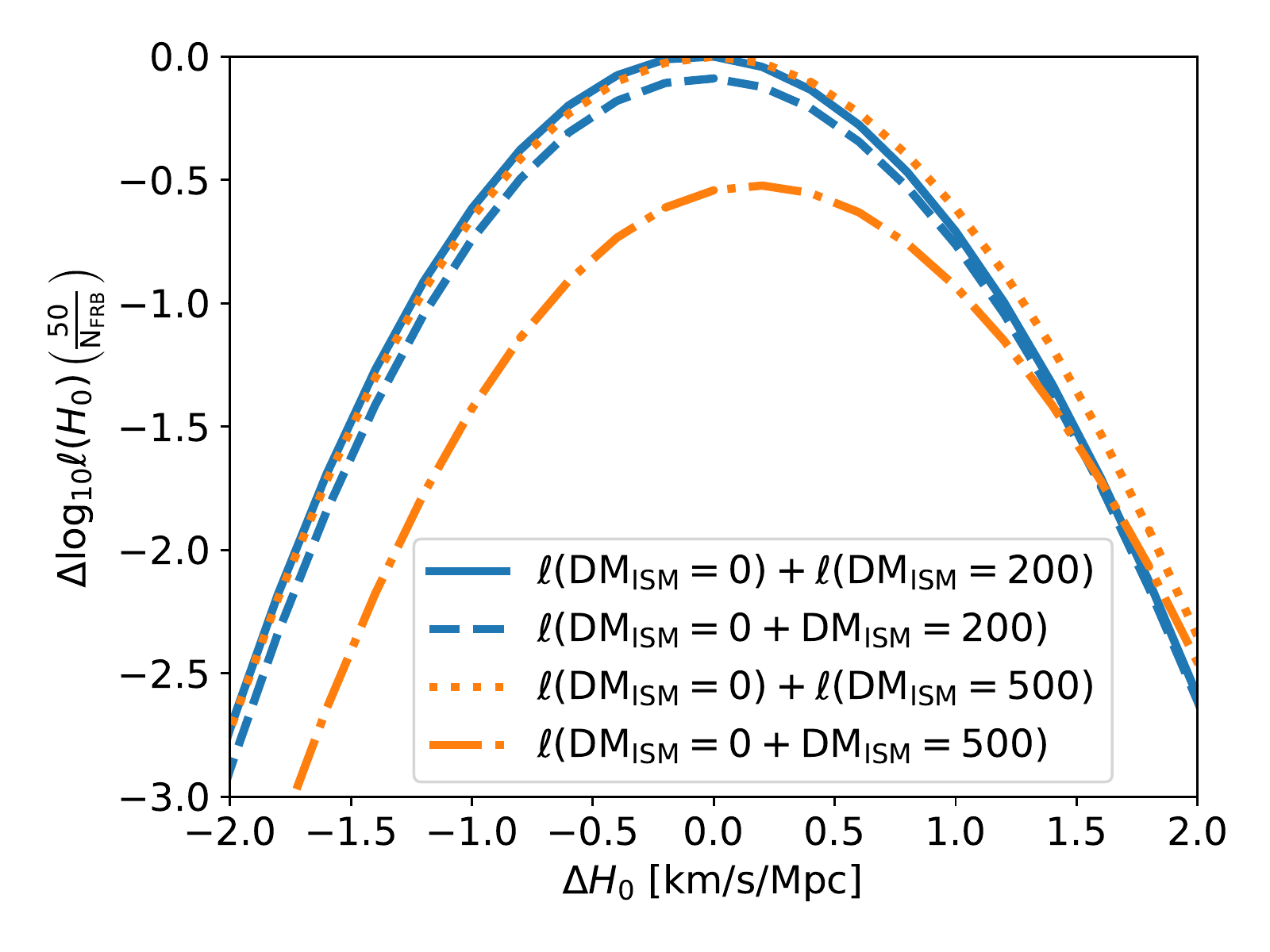}
    \caption{Effect of merging samples with common ${\rm DM}_{\rm ISM}$ on $H_0$.}
    \label{fig:dm_ism_bias}
\end{figure}

The effect of \dmism\ on \Hnot\ illustrated in \figref{fig:dm_ism_effect} is accounted-for in our simulation code, by subtracting \dmism\ (and \dmhost) from \dmfrb\ to calculate \dmeg\ on a per-event basis. However, the observation bias as a function of \dmeg\ is calculated using the mean value of \dmism\ for each sample only. This is done purely because generating a $z$--DM grid for each FRB individually is too computationally expensive.

To test the effect of mixing FRBs with different values of \dmism\ into the same evaluation, we take samples of FRBs generated at $\dmism=200$ and $500$ \pccc, and add them to the sample generated at $\dmism=0$ \pccc. \Hnot\ is then evaluated using the default method of averaging \dmism\ over the sample, i.e.\ to $\dmism=100$ and $\dmism=250$ \pccc\ respectively. This is compared to results on \Hnot\ generated by treating samples at $\dmism=0$, $200$, and $500$ \pccc\ individually, i.e.\ such that the average value of \dmism\ is the correct one. Results are shown in \figref{fig:dm_ism_bias}, giving the shift in most likely value of \Hnot, and the change in maximum likelihood $\ell_{\rm max}$ per 50 events (the vertical shift in maximum likelihood over 2000 events can be quite large, and prevents easy display). The effect of averaging \dmism\ over 0 and 200\,\pccc\ is negligible, resulting in bias of only +0.027\,\hubbunit. Mixing samples of FRBs with \dmism\ 
between 0 and 500\,\pccc\ is less reliable, with a bias of 0.18\,\hubbunit.

These biases are currently significantly smaller than the random errors from our small sample of localised FRBs, especially given that all have been detected with $\dmism < 200$\,\pccc. Thus we loosen the constraint of $\dmism<100$\,\pccc\ previously adopted in \citet{James2022Meth}, and include all FRBs localised by ASKAP in our main analysis. It also allows us to include ASKAP/FE and Parkes/Mb FRBs originally excluded from the sample of \citet{James2022Meth}, which are listed in Table~\ref{tab:parkes_frbs}.

\subsection{Effect of unlocalised ASKAP/ICS FRBs}
\label{sec:unlocalised}

The majority of ASKAP/ICS FRBs have had their host galaxies, and hence redshifts, identified. However, several FRBs remain unlocalised, as described in \secref{sec:no_hosts}. These reasons generally fall into two categories: reasons uncorrelated with FRB properties, which result in reduced statistical power on \Hnot\ but no potential bias; and those which are correlated with FRB properties, and thus could potentially bias a study such as that presented in this work.

In particular, the sample used in this work includes two FRBs with DM over 1000\,\pccc\ (FRB~20210407 and FRB~20210912) with potentially high-redshift host galaxies which have not been identified in initial follow-up observations with the VLT, although in the former case Galactic extinction plays a significant role.
If  \dmfrb\ is dominated by \dmcosmic\ therefore, the hosts of these FRBs will lie at $z \gtrsim 1$ and may not be detectable with standard follow-up observations. However, if \dmfrb\ is dominated by a large host contribution --- such as FRB~20210117A (\bhandariinprep)--- the host will be readily identified. This then leads to a sample biased towards low redshift.

\begin{figure}
    \centering
    \includegraphics[width=3.5in]{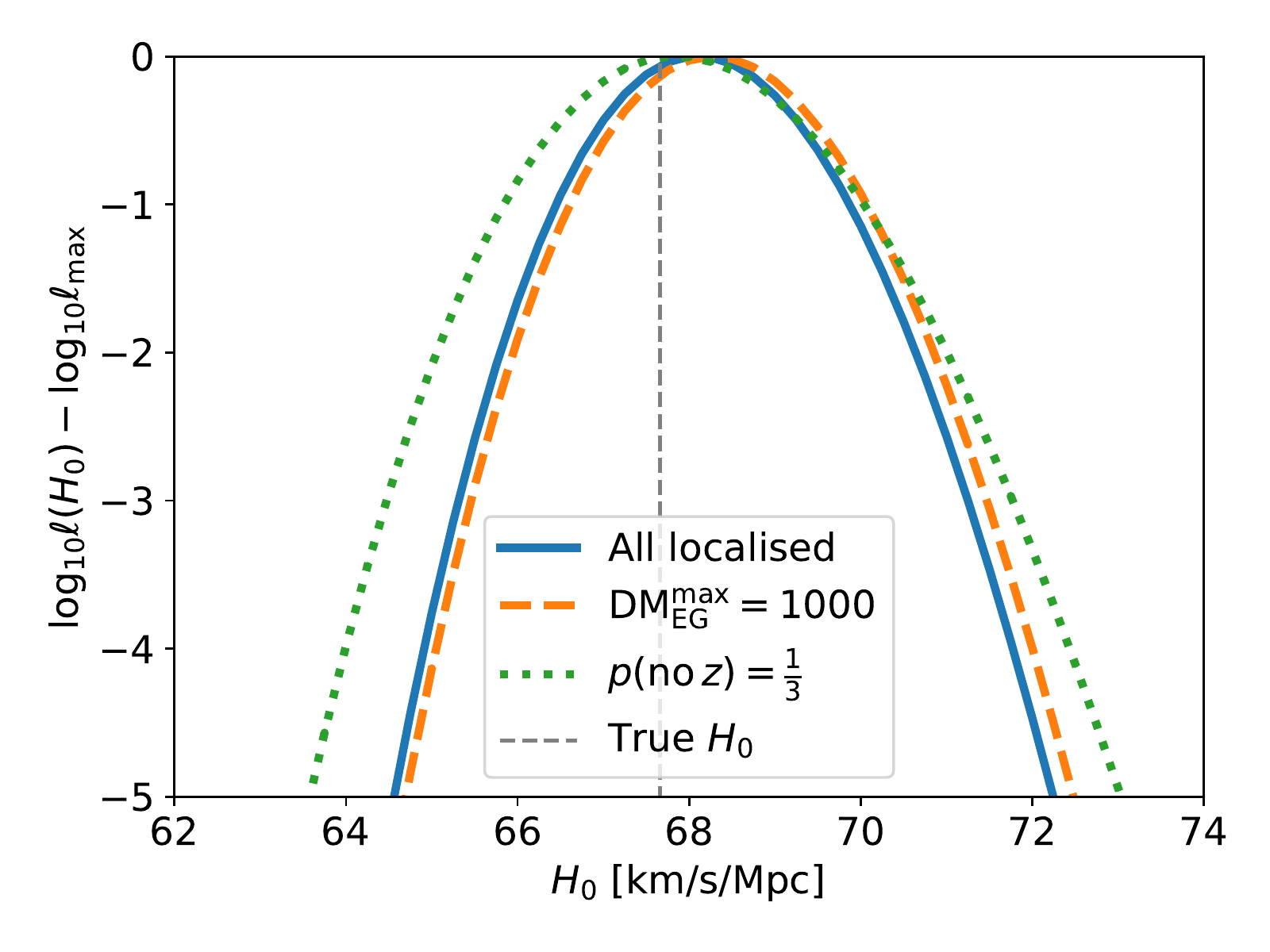}
    \caption{Constraints on \Hnot\ obtained from 1000 synthetic (Monte Carlo) FRBs when varying \Hnot\ only, assuming all bursts are localised; when discarding all 30 bursts with $\dmeg > {\rm DM}_{\rm EG}^{\rm max}=1000$ \pccc; and when also discarding one third of FRBs chosen randomly. No bias is visible when estimating \Hnot, but some loss of accuracy is, as expected.}
    \label{fig:missing_z}
\end{figure}

To counter this effect, we place an upper limit on \dmeg, $\dmeg^{\rm max}$, below which we expect all FRB hosts to be identifiable. Above $\dmeg^{\rm max}$, we discard all redshift information, even if known. We also include the DM of FRBs that remain unlocalised for observational reasons. This implementation has been tested using Monte Carlo FRBs generated from the \icsmid\ survey parameters, as per Appendix~\ref{sec:ISM_effect}. Results are shown in Figure~\ref{fig:missing_z}. Clearly, the loss of information when removing FRB redshifts results in poorer statistical constraints on \Hnot\ --- however, no systematic bias has been introduced.

\subsection{Effects of gridding in $z$--DM space}

\begin{figure}
    \centering
    \includegraphics[width=3in]{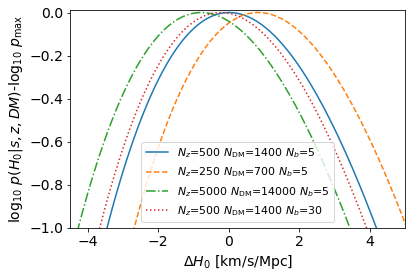}
    \caption{Simulated systematic effect on \Hnot\ estimates using the CRACO due to the number of redshifts $N_z$, DMs $N_{\rm DM}$, and beam values $N_b$. Shown is the posterior probability distribution $\log_{10} p(\Hnot|s,z,{\rm DM})$ normalised to the maximum value, as a function of $\Delta \Hnot$, i.e.\ the deviation from the best-fit value of \Hnot\ obtained for the default grid of $N_{\rm DM}$=1400, $N_z$= 500 and $N_b$=5.}
    \label{fig:zdm_grid}
\end{figure}

The probability distribution \pzdm\ is calculated on a finite grid of $N_z=500$ linearly spaced redshifts up to $z=5$ and $N_{\rm DM}=1400$ linearly spaced dispersion measures up to $\rm{DM}=7000$\,\pccc. Furthermore, we parameterise the telescope beamshape using only $N_b=5$ combinations of beam sensitivity $B$ and solid angle $\Omega(B)$. To test the effects of this choice of gridding, we again calculate $p(\Hnot|s,z,{\rm DM})$ for different gridding choices using the simulated CRACO sample while holding all other parameters fixed. The results are shown in \figref{fig:zdm_grid}.

We find that our best-fitting value of \Hnot\ varies by $\pm1$\hubbunit\ according to our choice of gridding: for this particular sample, a grid sparser by a factor of two ($N_z=250$, $N_{\rm DM}=700$) produces a higher value of \Hnot\ by 1\,\hubbunit, while ten-fold finer grid ($N_z=5000$, $N_{\rm DM}=14000$) produces a lower value of \Hnot\ by 1\,\hubbunit. Increasing $N_b$ has almost no effect on the estimation of \Hnot, which is likely due to this parameter being extensively optimised by \james.

We attribute this sensitivity to two regions of parameter space that are sensitive to smoothness, being the sharp decline in \pdmgz\ near the lower boundary due to \effect, and the low-$z$ region.

Compared to our current uncertainty in \Hnot\ estimates with FRBs, a potential systematic error of $\pm1$\,\hubbunit\ is small. However, in future, this suggests either using a finer grid in $z$--DM space; optimising the spacing, e.g.\ using log-spaced grids; or shifting to a method which does not use a brute force calculation over a grid, but performs a more intelligent optimisation of parameters.